\pdfoutput=1
\documentclass[
    aps,pre,twocolumn,
    reprint,
    superscriptaddress,
    nofootinbib,
    floatfix,
    amssymb,
    longbibliography
]{revtex4-2}
\usepackage[english]{babel}
\usepackage{amsmath}
\usepackage{amsfonts}
\usepackage{amssymb}
\usepackage[utf8]{inputenc}
\usepackage{nicefrac}
\usepackage[normalem]{ulem}
\usepackage{xcolor}

\definecolor{linkColor}{rgb}{0,0.3,0.7}
\usepackage[colorlinks=true,
            allcolors=linkColor,
            pdfborder={0 0 0},
            pdfencoding = auto
            ]{hyperref}

\usepackage{graphicx} 
\usepackage{dcolumn}  
\usepackage{bm}       
\usepackage{hyperref}
\usepackage[capitalise]{cleveref}

\usepackage[inline]{enumitem} 
\usepackage{multirow}

\newcommand{\K}{\ensuremath{\bar{K}}}
\newcommand{\n}{\ensuremath{n^*}}

\begin{document}

\title{Periodic temporal environmental variations \texorpdfstring{\\}{} induce coexistence in resource competition models}

\author{Tom Burkart}
\affiliation{Arnold Sommerfeld Center for Theoretical Physics and Center for NanoScience, Department of Physics, Ludwig-Maximilians-Universit\"at M\"unchen, Theresienstra\ss e 37, D-80333 M\"unchen, Germany}
\author{Jan Willeke}
\affiliation{Arnold Sommerfeld Center for Theoretical Physics and Center for NanoScience, Department of Physics, Ludwig-Maximilians-Universit\"at M\"unchen, Theresienstra\ss e 37, D-80333 M\"unchen, Germany}
\author{Erwin Frey}
\email{frey@lmu.de}
\affiliation{Arnold Sommerfeld Center for Theoretical Physics and Center for NanoScience, Department of Physics, Ludwig-Maximilians-Universit\"at M\"unchen, Theresienstra\ss e 37, D-80333 M\"unchen, Germany}
\affiliation{Max Planck School Matter to Life, Hofgartenstraße 8, D-80539 M\"unchen, Germany}

\date{\today}

\begin{abstract}%
Natural ecosystems, in particular on the microbial scale, are inhabited by a large number of species.
The population size of each species
is affected by interactions of individuals with each other and by spatial and temporal changes in environmental conditions, such as resource abundance.
Here, we use a generic population dynamics model to study how, and under what conditions, a periodic temporal environmental variation can alter an ecosystem's composition and biodiversity.
We demonstrate that using time scale separation allows one to qualitatively predict the long-term population dynamics of interacting species in varying environments.
We show that the notion of Tilman's \textit{R* rule}, a well-known principle that applies for constant environments, can be extended to periodically varying environments if the time scale of environmental changes (e.g., seasonal variations) is much faster than the time scale of population growth (doubling time in bacteria).
When these time scales are similar, our analysis shows that a varying environment deters the system from reaching a steady state, and stable coexistence between multiple species becomes possible.
Our results posit that biodiversity can in part be attributed to natural environmental variations. 
\end{abstract}
\maketitle

\section{Introduction}
\label{sec:introduction}%
In a healthy ecosystem, a wide variety of species coexist, interacting with each other through cooperative or competitive behavior~\cite{Schaeffer:1988uq,Rapport:1998wm}.
These interactions tend to be complex and entangled: the causal chain between a modification of the ecosystem and the corresponding observable effects, in particular on the biodiversity, are not always obvious~\cite{Chesson:2000,Jackson:2009bs,Saavedra.etal2017, Reese:2018ep}.
To describe how the population sizes of interacting species change over time for specific systems, various models have been developed (for a comparison, see, for example, Ref.~\cite{Zwietering:1990vh}).
Among those, the Monod model~\cite{Monod:1949wv} is presumably the most widely used model, as it includes the effect of growth-restricting parameters (usually a limiting resource) in the model.
When multiple species compete for such a limiting resource in a constant environment Tilman's R* rule~\cite{Tilman:1982rc} indicates which species will outcompete all others.

Only few models, however, account for the fact that the vast majority of ecosystems occurring in nature are subject to an external temporal structure~\cite{White.Hastings2020}, such as light availability during day-night cycles~\cite{Litchman:2001tc}, temperature variations during the change of the seasons~\cite{Ewing:2016ew}, or the circadian rhythm of vertebrate gut microbiota~\cite{Liang.FitzGerald2017},  despite accumulating evidence that organisms are strongly affected by such external temporal periodic variations.
For example, gene expression in the fungus \textit{Neurospora crassa} can be coupled to periodic temperature variations (entrainment)~\cite{Burt.etal2021}, and knock-out experiments on the bacterium \textit{Rhodopseudomonas palustris} identified a protein that enhances cell growth when exposed to light-dark cycles, but does not provide any advantage in constant environments~\cite{Ma.etal2016}.
On the level of microbial communities, it was shown that the composition of the mouse gut microbiome can be affected by an externally imposed day-night cycle~\cite{Thaiss.etal2014} or a time-restricted feeding schedule~\cite{Zarrinpar.etal2014,Mori:2017}. 
The strong statement of Tilman's R* rule does not hold anymore in variable environments, with Hutchinson's proposed solution to the ``paradox of the plankton'' as the most prominent counterexample~\cite{Hutchinson:1961,Brown:1989}.

\begin{figure}[!t]
\includegraphics[width = \columnwidth]{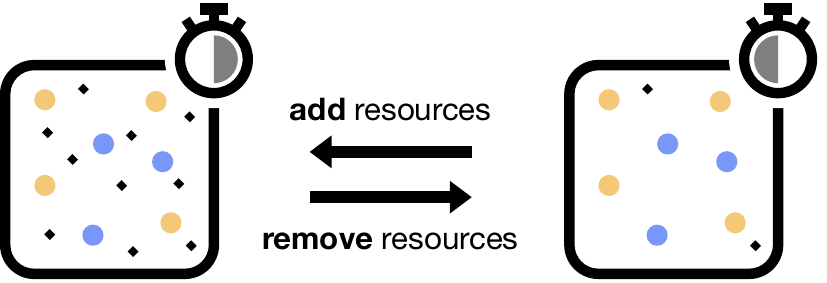}
\caption{(Color online)
Illustration of the resource competition model with periodically switching resource abundance. Yellow (or light gray) and blue (or gray) balls represent individuals of distinct species. Diamonds indicate available resource units.
For a fraction of each period, an abundant amount of resources are available (left).
For the remainder of the period, few or no resources are available (right), inhibiting the growth of the populations.
}
\label{fig:cartoon}
\end{figure}%

Theoretical models for population dynamics that consider external temporal structures mainly investigate the growth dynamics of a single population~\cite{Cushing:1986ut,Vance:1990tu,Baranyi:1993tx,Vance:1989tj,Coleman:1979wc,Hallam:1981wu,Pilyugin:2003ts,Nisbet:1976vd, Rosenblat:1980wr}, or study the interactions between multiple populations for fairly specific biological systems~\cite{Cushing:1980wp,White:2008bq,Levins:1979tu,Hale:1983ur,Litchman:2001tc,Namba:1984ut,Hsu:1980tf,Rosenblat:1980wr,deMottoni:1981ux,Gallagher:2020cb,Armstrong:1976,Hastings:1980, Posfai:2017}, with recently revived focus on stochastically varying environments~\cite{DeAngelis:1987, Yachi:1999, Maslov.Sneppen2017, Wienand.etal2018, Wienand.etal2017, Taitelbaum.etal2020, Pande.Shnerb2020, Yahalom.etal2019, Hufton.etal2019, Mancuso.etal2021, Ho.etal2022, Schreiber:2011, Hening.etal2021} and periodic resource cycles~\cite{Mori:2017, Erez.etal2020, Rodriguez-Verdugo.etal2019, Martinez.etal2022}.
These existing approaches show how populations change over time in specific systems;
however, as they are often tailored to a certain biological application, they cannot be applied in general to understand how interacting populations are affected by a varying environment, especially in terms of their long-term ability to coexist.
A notable exception is Chesson's widely used coexistence theory~\cite{Chesson:1981, Chesson:2000} which has also been extended to temporally varying environments~\cite{Chesson1994}.
Chesson's framework provides a rigorous mathematical analysis of ecological diversity and introduces useful concepts to theoretically assess multispecies coexistence.
However the rather technical formulation, revisions of theoretical concepts, and a lack of intuitive explanations for the analysis has lead to much confusion~\cite{Barabas.etal2018}.
Here, we aim to complement Chesson's foundations by providing an intuitive explanation and a physical perspective for a central question in the field:
How can a periodic temporal structure in general alter an ecosystem's ability to sustain or enhance biodiversity?
\par%

To address this question, we study a simple yet fairly general growth model that forbids coexistence in a constant environment according to the competitive exclusion principle~\cite{Armstrong:1980,Hardin:1960tg}.
We focus on the qualitative differences that appear when the environment is made explicitly time-dependent.
Our analysis reveals a mechanism by which the competitive exclusion can be overcome by periodically changing the environment for a system hosting multiple species [\cref{fig:cartoon}].
\par%

This paper is structured as follows:
We start in \cref{sec:model} by introducing a resource competition model, which is used as an example throughout this paper, along with a generalized growth model.
In \cref{sec:2spec}, we discuss the most important features of the generalized model and the implications of the competitive exclusion principle and Tilman's R* rule~\cite{Tilman:1982rc} in a time-independent environment.
We then study a system with two competing species that are subject to a periodic modulation of the environment, which is accounted for via a time-dependent resource abundance.
Finally, we extend the analysis to systems with more than two species, and study how a general external temporal periodic pattern impacts a more diverse ecosystem in \cref{sec:nspec}.
We conclude with a concise summary and an outlook.
\section{Model}\label{sec:model}
\subsection{The autonomous chemostat model}
Consider a population with $M$ distinct species, each with a population size $n_i(t)$, $i \in \{1,\ldots,M\}$.
Population growth is assumed to follow the Monod model~\cite{Monod:1949wv}, with a maximum per-capita growth rate~$\mu_i$.
All species, totalling a population size of ${N(t) = \sum_i n_i(t)}$, uniformly feed on and compete for a single common abiotic resource~$R$.
The term ``abiotic" refers to the resource abundance being constant or having an externally imposed time dependence, rather than being a dynamic quantity.
The resource is assumed to be replenished immediately after it is consumed (as a limiting case of a biotic resource, discussed in Appendix~\ref{app:chemostatModel}), such that the limiting factor for population growth is the amount of \emph{excess} resources ${R - N(t)}$.
Growth is only possible if there are excess resources available, which is made explicit by only allowing non-negative values of the excess resources.
The impact of resource scarcity is quantified by the species-dependent half-saturation constants~$K_i$~\cite{Armstrong:1980}.
Furthermore, the population size of each species is assumed to decrease at a per-capita rate~$\delta_i$.
Altogether, the dynamics of the population sizes are described by the following chemostat model~\cite{Grover:1997gr,Smith:1995tt}:
\begin{equation}
\label{eq:chemostat}
\frac{\mathrm{d}}{\mathrm{d} t} n_i(t) =
n_i(t) \cdot \left( \mu_i \, \frac{\text{max}(R - N(t),0)}{\text{max}(R - N(t),0) + K_i} - \delta_i \right).
\end{equation}
The choice of this chemostat model with an abiotic resource as a showcase is motivated by its simplicity and the clear intuition conveyed by each term.
Due to the limited resource availability, each species $i$ has a carrying capacity $\bar N_i$ at which the population growth is exactly levelled by decay.
This population size is given by ${\bar N_i = R - \K_i}$, where the offset $\K_i$ can be interpreted as a \emph{resource buffer}, denoting the amount of excess resources when species~$i$ reaches the steady state [\cref{fig:flow1}]; derivation in Appendix~\ref{app:resource_buffer}.
This resource buffer is conceptually similar to the equilibrium resource density $\text{R*}$~\cite{Tilman:1982rc}, in the sense that both quantities specify the equilibrium solution of the system.
Note that the index $i$ in the symbol $\bar N_i$ refers to the fact that the population $n_i$ can not grow if the total population size is larger than the carrying capacity of species $i$, i.e., if ${N > \bar N_i}$; a different species $j$ may still grow at this total population size if ${\bar N_j > N}$.
In the chemostat model, when a species does not face competition, this carrying capacity $\bar N_i$ is equal to the steady state population size $n^*_i$ (fixed point), and survival is only feasible for strictly positive carrying capacity.

Note that in Eq.~\eqref{eq:chemostat} we assume that all species consume the same amount of resources per capita, which can be derived from a more general version by appropriate rescaling as shown in Appendix~\ref{app:chemostatModel}.
One could further reduce the number of parameters by non-dimensionalizing the model, e.g., by expressing the time scale in terms of the death rates, however here we stick to the dimensional quantities to highlight the role of each of the parameters in the following analysis.

\subsection{The non-autonomous chemostat model}\label{sec:non-autonomous}
The model in Eq.~\eqref{eq:chemostat} describes an ecosystem that has no external temporal structures so that all model parameters remain constant in time, referred to as an \emph{autonomous} system.
Here, however, we are interested in a periodically varying environment (\emph{non-autonomous}), which we will incorporate via a time-dependent abiotic resource~$R(t)$.
In principle, any of the model parameters may depend on time, and one could also arrive at the following results using a time dependence other than the resource abundance; see for example Refs.~\cite{Smith:1995tt,Gallagher:2020cb,Miller:2017es, Taitelbaum.etal2020}.
However, it is more illustrative and biologically relevant~\cite{Litchman:2001tc,Burson:2018cf, Mori:2017} to study the case where the resource varies over time.

For simplicity, we assume that $R(t)$ switches periodically between two constant values $R_\text{a}$ and $R_\text{s}$, with a period duration of $T$, reminiscent of a seasonal cycle.
To further simplify the analysis, we explain the effect of an externally imposed time dependence for the case ${R_\text{s}=0}$ first, and generalize to ${0 \leq R_\text{s} < R_\text{a}}$ later.
The time window where resources are abundant (${R(t) = R_\text{a}}$) is assumed to last a proportion~$\nu T$ of the total period, with the \emph{activity ratio}~$\nu\in[0,1]$.
For the remainder of the total period, ${(1-\nu) \, T}$, resources are assumed to be scarce (${R(t) = R_\text{s}}$).
If no resources are available, then there should be no growth, which is enforced by only allowing non-negative values of the excess resources ${(R(t)-N(t))}$.
The resulting differential equation including the time-dependent resource~$R(t)$ reads
\begin{subequations}
\label{eq:td-model}
\begin{align}
\label{eq:td-chemostat}
\frac{\mathrm{d} n_i(t)}{\mathrm{d} t}  &=
 n_i(t) \left[ \mu_i \, \frac{\max\left(R(t) {-} N(t),0\right)}{\max\left(R(t) {-} N(t),0\right) + K_i} - \delta_i \right], \\
\label{eq:R-of-T}
R(t) &= 
\begin{cases}
	R_\text{a} & \text{for } 0 \,{\leq}\, t \,{<}\, \nu T \, , \\
	R_\text{s} & \text{for } \nu T \,{\leq}\, t \,{<}\, T \, ,
\end{cases}
\quad R(t{+}T) \,{=}\, R(t) \, .
\end{align}
\end{subequations}
This non-autonomous chemostat model will serve as an example to illustrate the results throughout this paper.
Within each time period $T$, there are therefore two distinct phases:
While resources are abundant, the populations grow just as in the autonomous chemostat model, \cref{eq:chemostat}.
This growth is impeded while resources are scarce, and in the special case ${R_\text{s}=0}$ that will be investigated first the population sizes decay exponentially.
Note that this is identical to a setup where populations are diluted by a constant factor periodically, as common in serial dilution experiments~\cite{Smith.2011, Gore.etal2009, Yurtsev.etal2016, Friedman:2017dk, Xie.Shou2021, Niehaus.etal2019}.

\subsection{General model class}
The chemostat model is a special case of a general class of growth models referred to as \emph{competing species models}~\cite{Smith:1986pc,Hirsch:1982so}.
In this class of models, the net per-capita growth rates are general growth functions $ f_i(\{n_j(t)\},\,t)$ for each species~$i$.
The term `competing' implies that the growth rates decrease for increasing population sizes, ${\partial_{n_j} f_i(\{n_j(t)\},\,t) \leq 0}$.

When the population growth depends on a linear combination of the individual population sizes $\sum_j q_j n_j$, the growth functions $f_i$ can be rewritten in terms of the total population size $N$.
This can be achieved by rescaling the population sizes by their relative weigthing factor $q_j$ as shown in Appendix~\ref{app:chemostatModel}.
The resulting class of models defined by
\begin{subequations}
\label{eq:general-model}
\begin{align}
\label{eq:competition-model}
&\frac{\mathrm{d}}{\mathrm{d} t} n_i(t)  =
n_i(t) \cdot  f_i(N(t),\,t) \, , \\
\label{eq:competing-condition}
&\frac{\mathrm{\partial}}{\mathrm{\partial} n_j} f_i(N(t),t)  \leq
 0
\end{align}
\end{subequations}
is a generalization of the non-autonomous chemostat model in Eq.~\eqref{eq:td-model}.
Additional constraints to make the growth functions $f_i$ realistic are stated in Appendix~\ref{app:growth_function}. 
We will show that the concepts that can lead to increased biodiversity in the resource competition model can actually be applied to the entire class of competing species models as specified in~\cref{eq:general-model}.

\goodbreak
\section{Two-species competition}
\label{sec:2spec}
\nobreak

In this section we begin the analysis of the role of time-dependent resources with a discussion of systems hosting two competing species.
For simplicity we assume that the decay rates $\delta_i$ are identical for both species, which is a basic feature of chemostat models where the decrease in each species' population size is mainly due to washout from the chemostat~\cite{Smith:1995tt}.
We will shortly review the R* rule in a constant environment before analyzing the time-dependent environment.
\begin{figure}[!tb]
\includegraphics[width=\columnwidth]{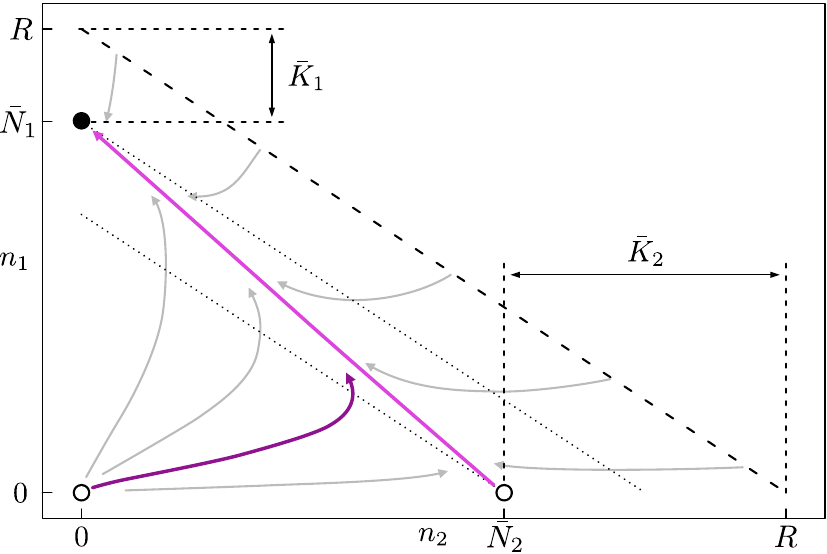}
\caption{\label{fig:flow1}%
(Color online)
Representative flow diagram showing the time evolution of the population sizes of two competing species.
The dashed line represents a subspace of constant total population size ${N=R}$, and the dotted lines represent subspaces of constant total population size $N=\bar N_1$ and $N=\bar N_2$.
Light gray arrows indicate the overall flow.
The pink (or gray) trajectory represents the heteroclinic orbit connecting the two non-trivial fixed points.
For two competing species, there are one stable (filled circle) and two unstable fixed points (open circles).
Following the R* rule, the stable fixed point corresponds to the species with the smallest resource buffer $\K_i$.
The purple (dark gray)  line represents a specific trajectory initially favoring the species with larger resource buffer, but finally resulting in the survival of the species with the smaller resource buffer, following the R* rule.}%
\end{figure}%
\par
\subsection{Competitive exclusion and the R* rule}
In a system hosting two or more distinct species whose population dynamics are described by \cref{eq:chemostat}, competition for a single limiting resource will allow only one survivor, namely the one with the smallest resource buffer $\K_i$.
This rule is known as the \emph{competitive exclusion principle}~\cite{Armstrong:1980,Hardin:1960tg}.
It can be understood heuristically in the following way: For two competing species, both populations can grow until the total population size $N$ approximately matches the carrying capacity $\bar N_i$ of one of the two species (purple or dark gray flow line in \cref{fig:flow1} approaching the heteroclinic orbit).
At this point, the net growth for this species is zero, whereas the other species (the one with the smaller resource buffer) can still grow, thereby further increasing the total population size.
Since this reduces the amount of resources available to the species with the larger resource buffer, its population size will start to decrease, eventually leading to extinction of the species with the larger resource buffer (flow along the heteroclinic orbit towards the stable fixed point in \cref{fig:flow1}).

In total, a system hosting two species therefore has three fixed points: assuming ${\K_1 < \K_2}$, the stable fixed point is located at ${(n_1, n_2) = (\bar N_1, 0)}$, whereas the two unstable fixed points are located at $(0, \bar N_2)$ and $(0,0)$.
The stable and unstable fixed points are represented as filled and open circles in Fig.~\ref{fig:flow1}, respectively.
A fine-tuned case of coexistence is possible when the two species have the same carrying capacity ${\bar N_1 = \bar N_2}$, where all population sizes ${n_2 = \bar N_1 - n_1}$ are neutral equilibrium solutions. 
If the system consists of more than two species, this recursive argument can be repeated to show that the species with the lowest resource buffer will be the only one to survive~\cite{Grover:1997gr}.

Generically, Tilman's R* rule--equivalent to finding the lowest equilibrium resource density $\text{R*}_i$ or the lowest resource buffer $\K_i$--predicts which species will survive when the total population size reaches the carrying capacity~\cite{Tilman:1982rc}.
Throughout our analysis, the population size of this surviving species (termed \emph{gleaner}, or K-strategist~\cite{Pianka:1970}) will be labeled by $n_1$.
While the gleaner species will dominate in the long run for high total population sizes, the other species (termed \emph{opportunist}, or r-strategist, $n_2$) may grow faster on short time scales but will be suppressed by the gleaner in the long run, as shown in \cref{fig:sampleTraj}a.
Such a scenario, known as \emph{gleaner-opportunist trade-off}~\cite{Grover:1997gr}, has been observed, for example, in phytoplankton competition~\cite{Hutchinson:1961, Litchman:2001tc, Barton.etal2010}.
For the chemostat model, Eq.~\eqref{eq:chemostat}, Tilman's R* rule implies that the most successful strategy for surviving in competition with other species is to minimize one's resource buffer~$\K_i$, i.e.~to optimize one's resource utilization.
Interestingly, this outcome solely depends on the resource buffer, but not explicitly on the growth rates of the populations~\cite{Smith:1995tt,Tilman:1982rc}.

\subsection{Reversal of survival}
\begin{figure}[t]
\centering
\scriptsize
\includegraphics[width = \columnwidth]{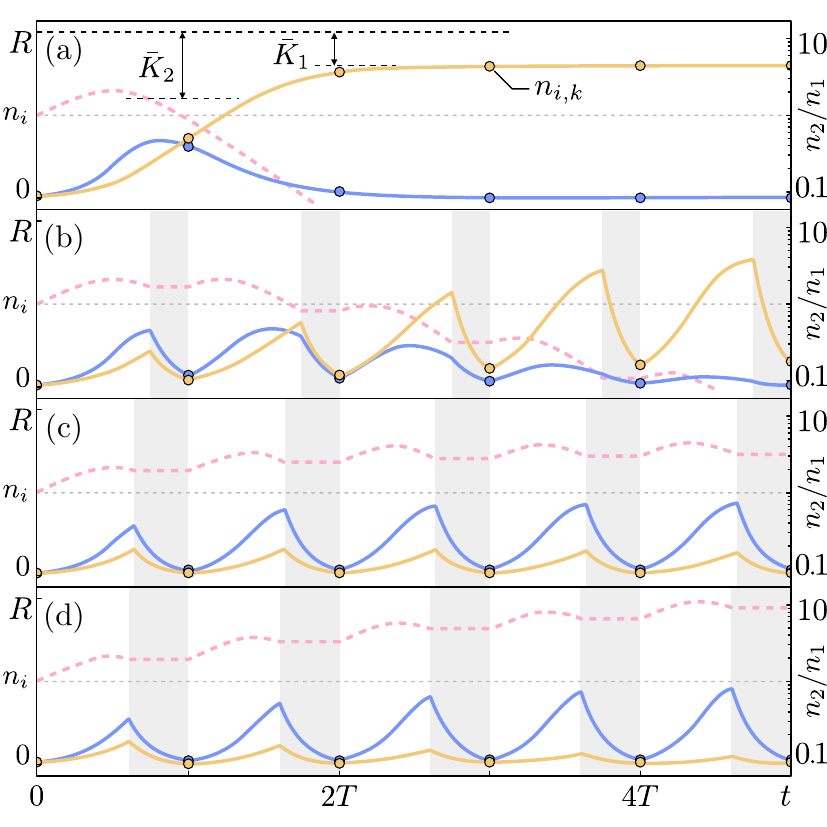}
\caption{\label{fig:sampleTraj}%
(Color online)
Comparison of the population dynamics at a fixed periodicity ${T = 7}$ for four different fractions $\nu T$ with resources available: (a) ${\nu = 1}$, (b) ${\nu = 0.75}$, (c) ${\nu \approx 0.64}$, and (d) ${\nu = 0.61}$ (remaining parameters in Table~\ref{tab:params}).
Gleaner ($n_1$) population sizes are shown in yellow (or light gray), the opportunist ($n_2$) is shown in blue (or gray).
Points indicate the discrete-time (between-season) population sizes $n_{i,k}$.
Dashed lines show the relative population size $n_2/n_1$ (log scale, right axis).
Shaded regions indicate time periods when resources are scarce.
In the time-independent case (a), the R* rule takes its full effect, with the opportunist approaching its steady state at ${R_\text{a}-\K_2}$ quickly, but being overtaken by the gleaner with the smaller resource buffer $\K_1$ eventually.
These dynamics are altered for time-dependent environments:
For increasingly long time episodes without resources, the gleaner's competitive advantage becomes less effective (b), eventually leading to coexistence at a neutral equilibrium (c), or even to inverted dynamics compared to the conventional R* rule (d).
Whether the external temporal periodic pattern leads to regular dynamics, inversion, or coexistence, depends on the duration of the period with resources absent, which is parametrized by the activity ratio $\nu$.}%
\end{figure}

In contrast to the time-independent case, the opportunist can gain a significant advantage from the quick growth at low total population sizes in an environment with periodically switching resources. This effect has been acknowledged before in the context of r- and K-specialists~\cite{Huston:1979} and will be summarized briefly in the following:
While resources are available and before the opportunist's population size can reach its maximum, the fast growth of the opportunist ensures that the relative population size $n_2(t)/n_1(t)$ increases to the advantage of the opportunist (dashed lines in Fig.~\ref{fig:sampleTraj}).
During the time period when resources are absent, both population sizes decrease, but the relative population size remains the same since we assumed equal decay rates $\delta_i$ here.
Note that, if this time period without resources is short enough, then the total population size remains high and the gleaner is not affected critically by the lack of resources, so that the opportunist loses to the gleaner in the long run [\cref{fig:sampleTraj}b].
In contrast, for a sufficiently long time period without resources, the populations are set back to very low population sizes each period.
Due to the fast growth of the opportunist at low total population sizes, the opportunist can take over in the long run, whereas the gleaner will go extinct, which is the \emph{inverse} result compared to the conventional R* rule [\cref{fig:sampleTraj}d]~\cite{Hutchinson:1961}.
At the transition from short to long time periods without resources, the opportunist's advantage at low population sizes and the disadvantage at high population sizes level each other, leading to \emph{coexistence} between the gleaner and the opportunist [\cref{fig:sampleTraj}c]. 
This concept of alternating periods with and without resources is intimately related to the ``storage effect'' that gives rise to coexistence in Chesson's coexistence theory~\cite{Warner:1985, Chesson:1986, Yachi:1999}.
Finally, when the period without resources is too long, all species may go extinct, thereby actually reducing the biodiversity of the ecosystem compared to an environment with constantly abundant resources.

\subsection{Bifurcation diagram: inversion \& coexistence}%
Under what conditions will the opportunist species survive instead of the gleaner species?
Heuristically, there are three conditions on the time dependence of the resource abundance that must be met for this reversal of survival to occur:
\begin{enumerate*}[label=(\roman*)]
 \item~The time window during which resources are available needs to be short enough so that competitive exclusion does not take effect.
 \item~The time window during which resources are absent needs to be long enough to ensure that the total population size decays to a small value before the resources become available again.
 \item~The time window during which resources are available needs to be long enough so that the populations can recover from the time window where resources are absent, as otherwise all populations would go extinct over the course of multiple periods, known as a storage effect~\cite{Warner:1985, Chesson:1986}.
\end{enumerate*}
These requirements impose constraints on the activity ratio $\nu$ that characterizes the external periodic structure.

In the following, we provide quantitative reasoning for these qualitative arguments.
To this end, we approximate the continuous dynamics with time-dependent resource abundance by a discrete equivalent (a map) with time-independent resource abundance.
Based on this approximation, we generalize the R* rule to time-dependent environments.
From this, we then derive the constraints on the activity ratio for reversal of survival and use an invasibility criterion~\cite{Davis.etal2005} to determine constraints on the period duration, validating the arguments above and allowing to estimate the bifurcation diagram for the long-term population dynamics.

\begin{figure*}[!htb]
 \centering%
 \scriptsize%
 \includegraphics[width = \textwidth]{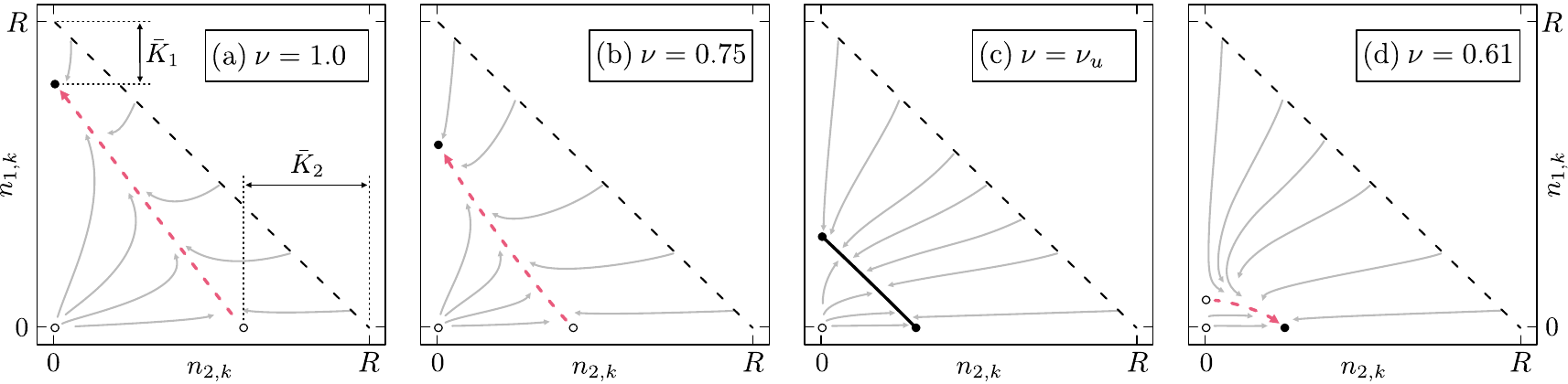}
 \caption{\label{fig:discrete_map}%
 (Color online)
 Flow diagrams of the discretized population dynamics $(n_{1,k},n_{2,k})$ obtained from the map Eq.~\eqref{eq:map} for the chemostat model at different values for $\nu$ as indicated in the graph; panel (c) shows the flow right at the upper activity ratio threshold $\nu_u \approx 0.64$.
 For $\nu = 1$, the flow diagram of the between-season population sizes (panel (a)) is identical to the continuous-time flow diagram shown in Fig.~\ref{fig:flow1}.
 Red (thick gray) dashed arrows represent heteroclinic orbits between the fixed points.
 At the threshold value $\nu_u$ for the activity ratio, all states on this orbit are stable (black line), corresponding to fine-tuned coexistence between the two species for a range of population size combinations ($n_1,\,n_2$), i.e., with neutral equilibria.
 For parameters see Table~\ref{tab:params}.
 }
\end{figure*}%

\subsubsection{Approximation of the population dynamics}
In the following, we study the population dynamics in the competing species model class with externally imposed time dependence, as specified in Eq.~\eqref{eq:general-model}.
Now, assume that nonlinear contributions to the population size changes $\partial_t n_i(t)$ within a single period $T$ can be neglected.
This assumption is reasonable if the population size remains approximately constant over the course of a single period $T$, meaning that ${n_i(t) \approx n_i(k \cdot T)}$ for ${t \in [k \, T, \, (k{+}1) \, T]}$.
In Appendix~\ref{app:map}, we show that this is valid for short period durations $T$ compared to the time scales of growth, ${T \ll 1/f_i}$. 

Using this approximation and writing ${n_{i,k} := n_i(k \cdot T)}$ and ${N_k := N(k \cdot T)}$ to denote the population size after $k$ periods (\emph{between-season} population size), one can approximate the continuous (\emph{within-season}) population dynamics in Eq.~\eqref{eq:general-model} by a discrete map:
\begin{align}
\label{eq:map}
    n_{i,k+1} \bigl(\{n_{j,k}\}\bigr)
   &= 
    n_{i,k} \cdot 
    \exp
    \biggl[ 
    \int_0^T \mathrm{d} t \, 
    f_i (N_k,t)
    \biggr]
    \nonumber \\
    &= n_{i,k} \cdot
    \exp
    \bigl[ T \cdot
    \langle  f_i( N_k ) \rangle 
    \bigr]
    .
\end{align}
Here, $ \langle  f_i( N_k) \rangle $ is the average growth rate over the course of one period $T$ at a given total population size $N_k$.
We provide a formal derivation of this map in Appendix~\ref{app:map}.
This approach has been used previously in Ref.~\cite{Vance:1990tu} to derive certain mathematical properties of a class of models including \cref{eq:general-model}, and it is similar to the averaging methods employed in Refs.~\cite{deMottoni:1981ux,Cushing:1986ut, Doulcier.etal2020}.
Note that this map is obtained assuming $T\ll 1/f_i$ and is merely a linear approximation of the actual dynamics, and we will discuss the nonlinear effects arising in the general case later on.

In the discrete map in Eq.~\eqref{eq:map}, the externally imposed time dependence (the resource abundance $R(t)$ in the chemostat model) is integrated out.
In other words, only the ``time-averaged" effect of the explicitly time-dependent parameters is of relevance for this approximation.
This greatly simplifies the qualitative analysis of the asymptotic dynamics of the population sizes at ${t\to \infty}$:
for the continuous system, the population dynamics approach a limit cycle [Fig.~\ref{fig:sampleTraj}c], but an analytic expression for the population sizes at the limit cycles can in general not be obtained due to the nonlinearity of the population dynamics.
In the discrete map, however, the population dynamics approach a fixed point $\{n_j^*\}$ which can be calculated from Eq.~\eqref{eq:map}.
This allows to analyse the system quantitatively while preserving the qualitative features.

Figure~\ref{fig:discrete_map} shows the time evolution in the form of a flow diagram for the population sizes $(n_{1,k},\ n_{2,k})$ as obtained from the discrete map Eq.~\eqref{eq:map}.
Note that these flow diagrams do not show the population size variations over the course of a single period, but rather the change of the population sizes over subsequent periods $\{k,\, k{+}1,\, \ldots\}$ (between-season).
Depending on the value of the activity ratio $\nu$, qualitatively different dynamics are observed:
In the absence of periodic changes of the environment (${\nu = 1}$), the gleaner species $n_1$ is the surviving species, as indicated by the stable fixed point in the flow diagram (filled circle at ${(n_{1,k}, n_{2,k}) = (\bar N_1, 0)}$).
This is the R* rule.
For a periodically changing environment with a sufficiently large activity ratio (${\nu > \nu_u}$, [Fig.~\ref{fig:discrete_map}b]), the gleaner remains the surviving species.
However, at a threshold value of the activity ratio, ${\nu = \nu_u}$, the two non-trivial fixed points are stable simultaneously, as well as all states $(n_1, n_2)$ along the heteroclinic orbit connecting these two fixed points.
This corresponds to coexistence between the two species, where the final state of the system depends on the initial conditions.
Note that this situation is similar to the fine-tuned coexistence of two species in a time-independent environment with $\bar N_1 = \bar N_2$ (neutral equilibria) but will expand into a coexistence region of finite size at longer period durations $T$ (stable equilibria).
For activity ratios below this threshold (${\nu < \nu_u}$, [Fig.~\ref{fig:discrete_map}d]), the opportunist is the surviving species.
Thus, upon changing the activity ratio $\nu$, the system exhibits a transition from a regime obeying the R* rule (gleaner survives, termed \emph{regular long-term dynamics} in the following), to a regime with inverted outcome compared to the R* rule (opportunist survives, termed \emph{inverted long-term dynamics}).
In the following, we will quantify this transition and the threshold value $\nu$ for the activity ratio by generalizing the R* rule to time-dependent environments.
To facilitate the comparison with the situation in a constant environment, the terms ``regular" and ``inverted" long-term dynamics take the dynamics in the time-independent environment as baseline throughout this analysis.
Finally, neither of the species can survive for too small activity ratios $\nu$ (equivalent to negative carrying capacities $\bar N_i$), which becomes obvious in the extreme case $\nu = 0$ where no resources are available at any time.

\subsubsection{Analogy to the R* rule}
Recall that in the autonomous system, the R* rule states that the species with the smallest resource buffer or, equivalently, the largest carrying capacity ${\bar N_i = R- \bar K_i}$ will survive.
If only one species prevails (${M = 1}$), this carrying capacity is identical to the steady-state population size $n^*_i$.
However, Tilman's R* rule holds only locally in time.
In the non-autonomous system, the resource buffer $\bar K_i$ is not well-defined due to the explicitly time-dependent growth rates $f_i$.
However, it is possible to infer a proxy for the resource buffer from the steady-state population size in a single-species system:
for the non-autonomous system, the analogue to each species' steady-state population size is the population size at the fixed point of the discrete map in Eq.~\eqref{eq:map},
$$n_{i,k+1} \bigl(\{\n_j\}\bigr) = n_{i, k} \, .$$
Note that when the between-season dynamics $n_{i,k}$ are at a fixed point the within-season dynamics $n_i(t)$ periodically vary over time [Fig.~\ref{fig:sampleTraj}c].
Following the line of arguments above in reverse, the fixed point in a system where only one species prevails (${M=1}$) corresponds to the largest sustainable population size, ${\bar N_i = \left. n^*_i \right|_{M=1}}$.
Equivalently, this allows to deduce a proxy resource buffer ${R_\text{a} - \left. n^*_i \right|_{M=1}}$ for the time-dependent system.
Based on this direct relation between the fixed point and the (proxy) resource buffer, we now generalize the R* rule to systems with externally imposed time dependence (within in the range of validity of the discrete map at $T \ll 1/f_i$):
\textit{In a system hosting two or more distinct species whose population dynamics are described by Eq.~\eqref{eq:general-model}, competition will allow only one survivor, namely the one with the largest maximally sustainable total population size ${\bar N_i \equiv \left. n^*_i \right|_{M=1}}$.}

\begin{figure*}[!htb]
 \includegraphics[width=\textwidth]{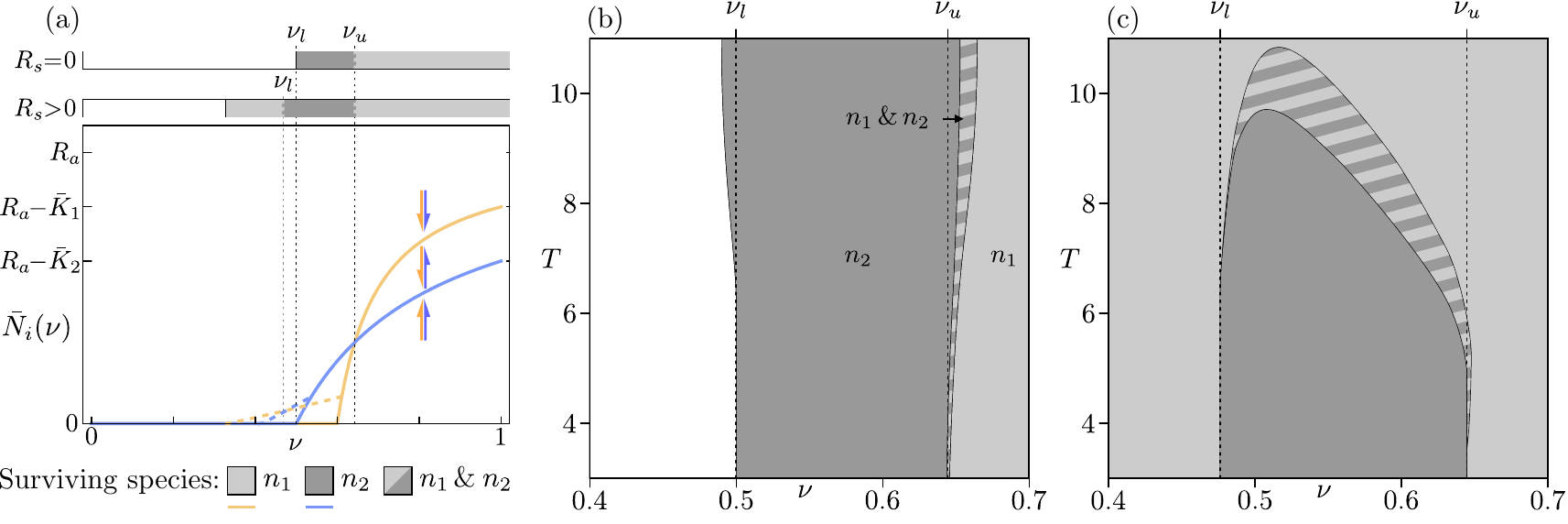}
 \caption{%
 \label{fig:phase_space}%
 (Color online)
 Qualitative competition dynamics.
 (a) Carrying capacities $\bar N_i(\nu)$ as a function of the activity ratio $\nu$ (analogue to fixed points of the growth functions $n^*_i$) for two distinct species (gleaner shown in yellow or light gray, opportunist shown in blue or dark gray) that exhibit inverted long-term dynamics for time-dependent resource abundance in the chemostat model.
 Solid lines correspond to zero resources in times of scarcity (${R_\text{s}=0}$), dashed lines show the fixed points for small but non-zero resources in this time window (${R_\text{s}>0}$).
 Vertical dashed lines indicate the calculated upper and lower boundaries $\nu_u$ and $\nu_l$ to the region of inverted long-term dynamics.
 Arrows indicate the flow of the system towards the nullclines.
 Bars above indicate the surviving species over all values of the activity ratio $\nu$ for ${T \to 0}$ using ${R_\text{s} = 0}$ and ${R_\text{s} > 0}$, respectively.
 Shaded/patterned regions indicate which species survives.
 In this and all other figures showing bifurcation plots, uniformly shaded areas correspond to a single species surviving the competition, while striped areas correspond to coexistence between two species (c.f.~legends below panel (a)).
 White areas indicate regions where neither species survives.
 (b) Representative bifurcation diagram for the chemostat model, which has no resources during episodes of scarcity (${R_\text{s}=0}$).
 For short period durations $T$, the calculated phase boundaries $\nu_u$ and $\nu_l$ (dashed lines) agree well with the data (solid phase boundaries).
 Deviations at large $T$ are due to nonlinearities in the population dynamics.
 (c) Representative bifurcation diagram for the chemostat model, which has small but finite resources during episodes of scarcity (${R_\text{s}>0}$).
 Again, the calculated phase boundaries (black dashed lines) match the data (solid phase boundaries) at short $T$.
 As predicted by Eq.~\eqref{eq:app:threshold_period_duration}, the conventional R* rule holds at large $T$ for any value of the activity ratio $\nu$. At intermediate $T$, a band of coexistence connects the regions of regular and inverted long-term dynamics.
 For (b) and (c), the data was obtained as described in Appendix~\ref{app:simulation} and the phase boundaries were interpolated.}
\end{figure*}

The fixed point values -- and thus also the carrying capacities $\bar N_i$ -- depend on the details of the external time dependence, which is effectively accounted for by the time-averaged growth rates $\langle f_i \rangle$ in the discrete map.
In particular, they depend on the activity ratio, $\bar N_i(\nu)$, shown in Fig.~\ref{fig:phase_space}a for two distinct species obeying the chemostat model.
In particular, the results from the autonomous system are recovered for ${\nu = 1}$, where ${\bar N_i (1) = R - \bar K_i}$ [Fig.~\ref{fig:discrete_map}a].
For the non-autonomous system, the predicted survivor is indicated by the bars above the plot frame in Fig.~\ref{fig:phase_space}a.
So far, only the case ${R_\text{s} = 0}$ (top row, solid lines) has been discussed; a generalization to ${R_\text{s} > 0}$ will follow in Section~\ref{sec:generalization}.

\subsubsection{Parameter regime for reversal of Tilman's R* rule}%
Following the above generalisation of the R* rule, the transition from regular to inverted long-term dynamics in a two-species system is found by determining the threshold value~$\nu_u$ for which both species have the same carrying capacity, ${\bar N_1 (\nu_u) = \bar N_2(\nu_u)}$.
Graphically, $\nu_u$ can be determined from the intersection of the functions $\bar N_i(\nu)$ [Fig.~\ref{fig:phase_space}a].
Since the carrying capacity is derived from the single-species fixed points $\left. n^*_i \right|_{M=1}$, one can use the discrete map in Eq.~\eqref{eq:map} to derive this threshold activity ratio.
In particular, this requires to find the parameter combination $(N, \nu)$ at which both species' averaged growth rates are zero, i.e., solving the system of equations
\begin{subequations}
\label{eq:nu-cond}
\begin{align}
    0 &= \langle \left. f_1 ( \bar N) \right|_{\nu_u} \rangle = \frac{1}{T} \int_0^T \mathrm{d} t \, \left. f_1\left(\bar N,t\right) \right|_{\nu_u} \\
    0 &= \langle \left. f_2 ( \bar N) \right|_{\nu_u} \rangle
\end{align}
\end{subequations}
for $\bar N$ and $\nu_u$.
For the chemostat model with zero resources during the period of scarcity (${R_\text{s} = 0}$), this system of equations has only one physical solution, which we derive explicitly in Appendix~\ref{app:phase_boundaries}.
Hence, in this case there is only one threshold activity ratio for which fine-tuned coexistence is possible [Fig.~\ref{fig:phase_space}a], separating the regions of regular and inverted long-term dynamics.
The region of inverted long-term dynamics is bound from below by another threshold activity ratio $\nu_l$, at which the overall resource abundance is too low to sustain even the opportunist population and both species go extinct.
Thus, the externally imposed time dependence can lead to three distinct characteristic outcomes in the chemostat model with ${R_\text{s}=0}$: regular long-term dynamics, inverse long-term dynamics (enclosed by upper and lower boundaries $\nu_u$ and $\nu_l$), and a system where neither species survives. 
In the limit of short period times T and $R_\text{s}=0$ in the chemostat model the phase boundaries take the values
\begin{subequations}
\begin{align}
    \nu_l &= \frac{\delta_2}{\mu_2} \cdot \frac{K_2 + R_\text{a}}{R_\text{a}} \,,\\
    \nu_u &= \frac{\delta_1 \delta_2 (K_1 - K_2)}{\mu_2 \delta_1 K_1 - \mu_1 \delta_2 K_2} \, 
\end{align}
\end{subequations}
as derived in Appendix~\ref{app:phase_boundaries}.

In general, there can be multiple threshold activity ratios (dashed lines in Fig.~\ref{fig:phase_space}a for ${R_\text{s} > 0}$), which can give rise to rich phase diagram structures.
In any case, the region of inverted long-term dynamics extends only over a fraction of the entire parameter space, i.e., ${0 < \nu_{u,l} < 1}$ are strict bounds.
This is evident since the cases ${\nu = 1}$ and ${\nu = 0}$ formally correspond to time-independent systems, for which Tilman's R* rule holds.
Thus, if an upper boundary $\nu_u$ for the region of inverted long-term dynamics exists, then there is also a lower boundary $\nu_l$; however, the nature of this boundary can vary.
Either, both species go extinct for ${\nu < \nu_l}$ (as in the case ${R_\text{s} = 0}$ in the chemostat model), or there is another region of regular long-term dynamics (possible for ${R_\text{s} > 0}$).

The discrete map in Eq.~\eqref{eq:map} is based on the assumption that the period duration is much smaller than the time scales of growth, ${T \ll 1/f_i}$, and thus all ensuing predictions are expected to hold only in this parameter regime in general.
In the chemostat model with ${R_\text{s}=0}$, however, numeric solutions of the system show that the predictions remain approximately valid for all period durations.
A sample phase diagram for the two-species chemostat model is shown in Fig.~\ref{fig:phase_space}b.
The vertical dashed lines indicate the phase boundaries as predicted by the discrete map, which agree well with the numerically determined phase boundaries for short period durations.
However, for large period durations, slight deviations from the predicted phase boundaries emerge: the region of inverted long-term dynamics (dark gray, labelled ``$n_2$'') broadens marginally.
In addition, the fine-tuned neutral-equilibrium coexistence at the upper boundary $\nu_u$ turns into a narrow band of stable coexistence (striped).

All significant deviations from the predicted phase boundaries happen within the parameter regime shown in Fig.~\ref{fig:phase_space}b.
In particular, the widths of the regions of inverted long-term dynamics and coexistence do not change anymore for even larger $T$.
Note that the approximate validity of the predicted phase boundaries for all $T$ is a special feature of the chemostat model.
In essence, this is due to the fact that there is no competition during the period of scarcity if ${R_\text{s} = 0}$ (without resources, there is no resource competition) and that for a significant portion of the period of abundance the population sizes are small (nonlinearities are negligible).
As a consequence, the effects of the nonlinearities that were neglected in the discrete map are mostly suppressed for all $T$ (since linearization along the lines of Eq.~\ref{eq:map} is possible for small population sizes), resulting in only minimal changes to the phase diagram structure for large period durations.
In the following, we will turn towards a more general case and study the case where the nonlinearities are not suppressed.

\subsection{Generalization to non-zero resources}
\label{sec:generalization}
So far, we have limited the analysis to ecosystems where no resources are available at all during the time window when resources are scarce.
Now, we relax this restriction and allow a limited, but non-zero, amount of resources ${0 < R_\text{s} < R_\text{a}}$.
In this case, a finite amount of resources is always available, and therefore there is always resource competition between the two species.
Thus, in contrast to the case ${R_\text{s}=0}$, the nonlinearities are now no longer suppressed.
More precisely, nonlinear competition dominates the population dynamics once the total population size drops below the resource abundance, ${N(t) < R(t)}$.
Since the gleaner population is less susceptible to resource scarcity and is less negatively affected by the competition than the opportunist, this implies that the gleaner gains an advantage during the period with scarce resources.
This effect counteracts the advantage that the opportunist gains from a quickly varying environment.
Thus, inverted long-term dynamics as discussed above is not necessarily established if the total population size drops below the threshold value $R_\text{s}$.

Under what conditions does the total population size drop below this threshold, ${N(t) < R_\text{s}}$?
This is the case only when the absolute duration of the periods with scarce resources is long compared to the time scales of growth, $1/f_i$:
for short period durations ${T \ll 1/f_i}$, the population sizes change only marginally during one cycle of the external oscillation, so that the total population size does not cross the threshold value $R_\text{s}$.
Consequently, the previously derived results (for ${R_\text{s}=0}$) and phase boundaries still apply in this more general case for ${T \ll 1/f_i}$: for a range of activity ratios $\nu$, there are inverted long-term dynamics (dark gray region in Fig.~\ref{fig:phase_space}c).
In contrast, the population sizes change significantly for long period durations ${T \gg 1/f_i}$, so that the total population size can cross the threshold $R_\text{s}$ during the time window of scarce resources.
As discussed above, this means that the gleaner species gains an advantage, so that regular long-term dynamics are restored in this parameter range.
At intermediate durations ${T\sim 1/f_i}$ the short-term advantage for the opportunist (within ${\nu_l<\nu<\nu_u}$) and the long-term advantage for the gleaner level each other.
This gives rise to intricate nonlinear interactions between the two species that can lead to coexistence over a finite parameter region as discussed in Appendix~\ref{app:phase_boundaries}.

In short, this coexistence can be understood as a case of mutual invasibility~\cite{Grainger.etal2019}.
A species $i$ can invade the other species $j\neq i$ if its average net growth over a single period is positive,
\begin{equation}
    \label{eq:invasibility}
    \langle f_i \rangle \left (\nu,\, T) \right|_{N = \n_j(t)} = \frac{1}{T} \int\limits_0^T \mathrm d t \, f_i(\n_j(t),\, t) > 0 \,.
\end{equation}
Since the within-season dynamics of the resident species $\n_j(t)$ also depend on $T$, the average net growth rates $\langle f_i \rangle(T)$ are nonlinear functions of the period duration $T$ and may both be positive for a range of $T$, corresponding to coexistence with a stable equilibrium solution.
The three cases (${T \ll 1/f_i}$, ${T\sim 1/f_i}$, and ${T \gg 1/f_i}$) coincide with those proposed by Hutchinson when discussing the surprising biodiversity of phytoklankton~\cite{Hutchinson:1961}.
Such a dependence of the coexistence of two species on the period duration was also recently quantified experimentally~\cite{Rodriguez-Verdugo.etal2019, Martinez.etal2022}.

Figure~\ref{fig:phase_space}c shows a sample phase diagram for a two-species chemostat model with ${R_\text{s} > 0}$.
The predictions from the discrete map in Eq.~\eqref{eq:nu-cond} (in particular, the carrying capacities $\bar N_i$) for this specific system are shown in Fig.~\ref{fig:phase_space}a as dashed lines.
The phase diagram differs from the phase diagram corresponding to $R_\text{s} = 0$ in three aspects.
First, the lower phase boundary $\nu_l$ is shifted as a result of the nonlinear resource competition.
Second, since the gleaner can now feed on the small amount of resources during the period of scarcity, it can survive for a range of activity ratios \emph{smaller} than the lower phase boundary $\nu_l$, instead of going extinct (light gray region).
Third, the region of inverted long-term dynamics does not extend to $T \to \infty$, but is capped by a band of coexistence instead (striped).
From the invasibility criterion Eq.~\eqref{eq:invasibility} one may find period durations $T_i(\nu)$ such that $\left.\langle f_i \rangle  \right|_{T_i}=0$ which correspond to the boundaries of the coexistence region.
A crude but intuitive estimate for these $T_i$ are the time scales of growth $1/f_i$, which we discuss further in Appendix~\ref{app:phase_boundaries}.

Note that even though the periodic variation of the environment is responsible for the coexistence it might be misleading to think of the two distinct states ($R(t) = R_\text{a}$ and $R(t) = R_\text{s}$) as dedicated ``niches'' for the gleaner and the opportunist, respectively.
In fact, the gleaner can have a competitive advantage in both states, as becomes evident in the limit $T\gg 1/f_i$ [Fig.~\ref{fig:phase_space}c].
Instead, the temporal niche during which the opportunist can thrive is determined by the environmental variation in combination with the gleaner's population dynamics: any time period where the total population size is much smaller than the opportunist's carrying capacity, $N(t) \ll \bar N_2$, is a temporal niche for the opportunist.
Similarly, any period where the total population size is larger than the opportunist's carrying capacity,  $N(t) > \bar N_2$, is a temporal niche for the gleaner.
The niches are therefore self-shaped by the ecosystem, and introducing additional species to the ecosystem can create additional niches allowing for richer biodiversity and coexistence between many species.

\section{Many-species competition}\label{sec:nspec}
Until now, we explained how coexistence and inverted long-term dynamics can arise in a time-dependent environment hosting two species.
However, natural ecosystems outside laboratory conditions typically consist of more than two species.
In a time-independent environment, the competitive exclusion principle holds for an arbitrary number of speciesas long as the resource consumption is not constrained~\cite{Armstrong:1980,Hardin:1960tg, Posfai:2017}.
In contrast, it is known that competitive exclusion can be overcome for resource competition in time-dependent environments, e.g.\ by successive temporal niches~\cite{Armstrong:1976,Chesson:2000}, by introducing biotic resources~\cite{Koch:1974,Tilman:1982rc,White:2008bq}, or both~\cite{Levins:1979tu,Nowack:2018}.
However, it is not clear in general how multiple competing species with complex interaction networks can coexist in a periodically varying environment.
In the following, we demonstrate how the qualitative dynamics of such complex networks can be assessed by means of the theoretical framework that we used to analyze the two-species competition in Section~\ref{sec:2spec}.
\par

In order to understand the population dynamics of $M$ competing species ($M$-species competition), we use that in competing species models the pairwise interactions between a set of species provide qualitative information about the interactions involving all community members (non-pairwise)~\cite{Klausmeier:2010hp,Faust:2012kq,Friedman:2017dk}.
While it is known that many-species interactions can enhance the stability of biodiverse ecosystems and cannot be disregarded in general~\cite{Bairey.etal2016, Levine.etal2017} the restriction to pairwise interactions is sufficient for the models in Eq.~\eqref{eq:general-model} where only the total population size is relevant for the competition;
the validity of this statement is shown in Appendix~\ref{app:pairwise}.
Heuristically, this follows from the fact that the relative impact of a species on the competition decreases as its relative population size, $n_i / N$, decreases.
Consequently, the species with the largest population size has dominant impact on the competition.
Other species therefore mainly compete with this dominant population, and competition between two comparably small populations has only negligible impact on the overall dynamics.
Combining the insights from all pairwise interactions, this allows to characterize the phase diagram for many-species competition.
In the following we explicitly demonstrate this for three-species competition.

\subsection{Bifurcation diagram}
\begin{figure*}[!t]
    \centering
    \includegraphics[width = \textwidth]{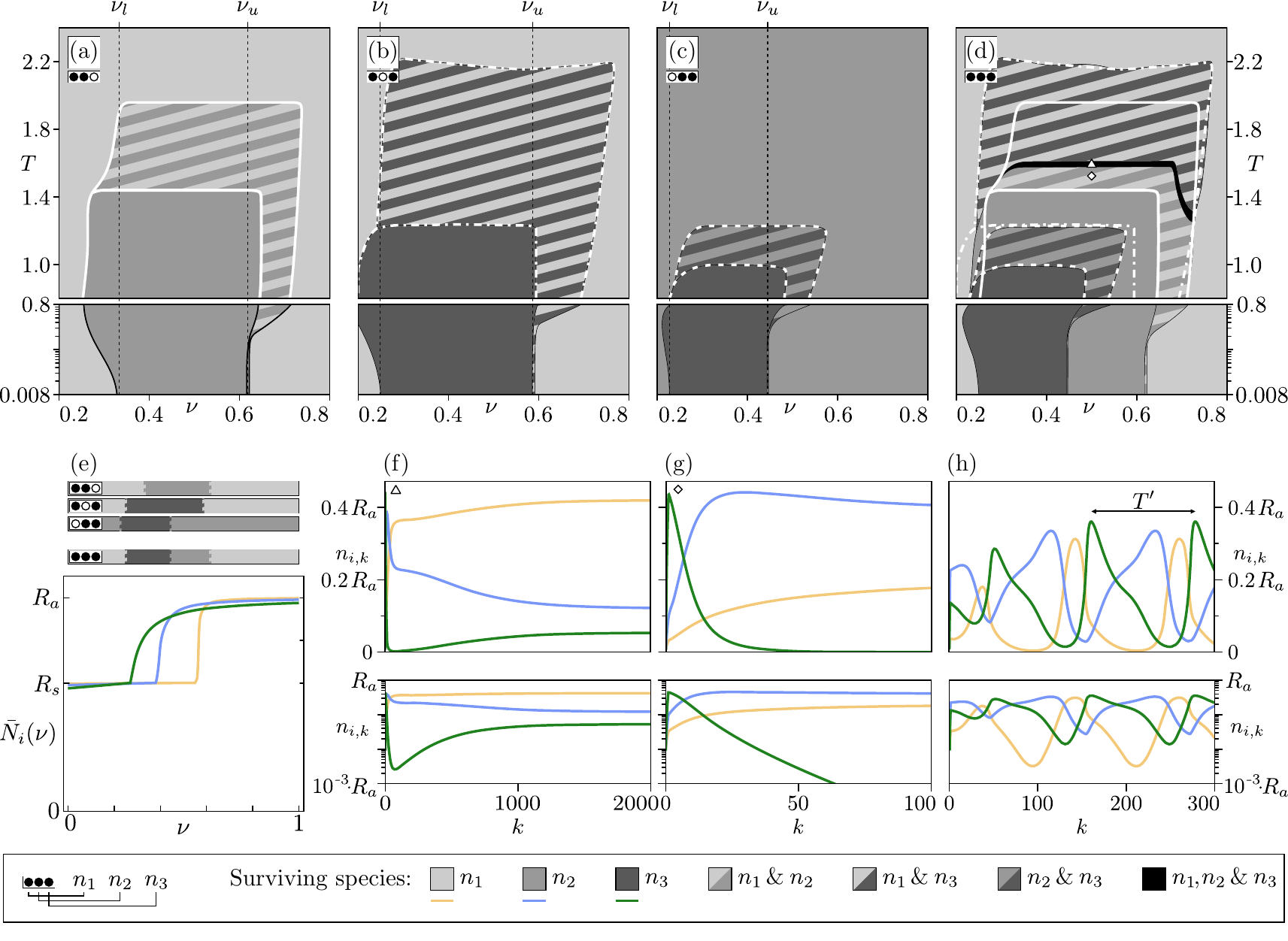}
    \caption{\label{fig:3specPhaseDiagram}%
    (Color online)
    Three-species competition. 
    For each bifurcation diagram, filled (open) circles indicate that the corresponding species takes part (does not take part) in the competition.
    (a-c) Phase diagrams for pairwise competition with parameters as specified in Table~\ref{tab:params}, with $n_1$, $n_2$ and $n_3$ represented by light, medium, and dark gray shading, respectively.
    White lines indicate phase boundaries for reference in panel (d).
    For the chosen parameter set, each pair of species exhibits a region of inverted long-term dynamics and coexistence.
    The (vertical) dashed black lines indicate the calculated phase boundaries for the limit ${T\to0}$.
    (d) Phase diagram for the competition of all three species. The two-species phase boundaries (white lines) are copied from the pairwise phase diagrams and show good agreement with the actual phase boundaries for the three-species system. The phases for three-species competition can be predicted from combining the pairwise phase diagrams.
    (e) Fixed points of the growth functions for the species shown in (a-d), exhibiting overlapping parameter ranges for the activity ratio $\nu$ with inverted long-term dynamics.
    Bars above the fixed points indicate the surviving species as predicted by the linearized map for ${T\to0}$.
    (f, g) Sample trajectories for the between-season population sizes $n_{i,k}$ for three-species competition, with coexistence between all three species (f) and pairwise $n_1$-$n_2$-coexistence (g), on a linear scale (top) and log scale (bottom) for better visibility.
    The corresponding values of $\nu$ and $T$ for these trajectories are marked in the three-species bifurcation diagram as triangle (f) and diamond (g).
    (h) Sample trajectory for the between-season population sizes $n_{i,k}$ for three-species competition with ${\delta_i \neq \delta_j}$, showing limit cycle dynamics with periodicity $T' \approx 120 T$ instead of approaching a fixed point.
    The corresponding parameters are specified in Table~\ref{tab:params}.
    }
\end{figure*}
Consider an ecosystem hosting three different species, each obeying~\cref{eq:td-model}.
The species are ranked from smallest to largest resource buffer in the following, so that ${\K_1 < \K_2 < \K_3}$ (full parameter set in Table~\ref{tab:params}).
With constant resource abundance only the species with the lowest resource buffer can survive as consequence of Tilman's R* rule ($n_1$ with the specified ranking).
The two other species will go extinct successively: first, the species with the largest resource buffer ($n_3$), and afterwards the species with moderate resource buffer ($n_2$).
In terms of pairwise interaction, the moderate species~$n_2$ takes the role of the opportunist when competing with~$n_1$, and it takes the role of the gleaner when competing with~$n_3$.
For all pairwise interactions, the species with highest (lowest) resource buffer always takes the role of the opportunist (gleaner).
\par
When the resource abundance switches periodically, the relative competitive advantages between pairs of species can be altered, in accordance with the observations in \cref{sec:2spec}:
for each pair of species, there may exist a range of activity ratios $\nu$ and period durations $T$ for which Tilman's R* rule fails, and instead the pair shows inverted long-term dynamics or coexistence [Fig.~\ref{fig:phase_space}].
In the three-species chemostat model with equal washout rates $\delta$, coexistence between all three species due to the time-dependent environment can be achieved when at least two out of the three pairs of species establish pairwise coexistence.

The most suitable representation for $M$-species competition is a phase diagram, which can be inferred from the set of all two-species phase diagrams.
The two-species diagrams for the sample system specified in Table~\ref{tab:params} are shown in Figs.~\ref{fig:3specPhaseDiagram}a-c, which were each obtained by eliminating one of the three species from the system and then solving the corresponding ODEs numerically.
The characteristics of these phase diagrams can be calculated analytically as shown in Section~\ref{sec:2spec}.
Notably, each of the three subsystems shows regions of inverted long-term dynamics and coexistence.
The phase diagram for three-species coexistence is shown in Fig.~\ref{fig:3specPhaseDiagram}d overlaid with the phase boundaries from all pairwise competitions.

The phase boundaries from pairwise competition predominantly match the three-species diagram, owed to the fact that pairwise competition is sufficient to explain $M$-species competition.
For the specific system shown in Fig.~\ref{fig:3specPhaseDiagram}, the population $n_3$ outcompetes the other two populations individually in the parameter range ${0.8 \lesssim T \lesssim 1.0}$ and ${0.25 \lesssim \nu \lesssim 0.45}$ (dark gray regions in Fig.~\ref{fig:3specPhaseDiagram}c, which is also enclosed by the dark gray region in Fig.~\ref{fig:3specPhaseDiagram}b).
In this parameter range, the population $n_2$ outcompetes $n_1$ (medium gray region in Fig.~\ref{fig:3specPhaseDiagram}a).
Therefore, in three-species competition, the population $n_1$ is outcompeted jointly by $n_2$ and $n_3$, and subsequently the latter outcompetes the former, so that only $n_3$ survives in the specified parameter regime (dark gray region in Fig.~\ref{fig:3specPhaseDiagram}d).
Similarly, the overall gleaner $n_1$ is defeated by $n_2$ and $n_3$ independently in pairwise competition across the entire parameter range for which coexistence between $n_2$ and $n_3$ is possible (striped region in Fig.~\ref{fig:3specPhaseDiagram}c).
Thus, in three-species competition, the overall gleaner $n_1$ cannot survive in this parameter region either.
Consequently, the phase boundaries from pairwise competition between $n_2$ and $n_3$ are valid for the three-species competition in this parameter region, too (same region in Fig.~\ref{fig:3specPhaseDiagram}d).
These rules for inferring the surviving species in $M$-species competition apply for the almost the entire parameter space, with exceptions where regions of pairwise coexistence overlap (see below).
To demonstrate this for the specific example shown here, the phase boundaries from pairwise competition [Fig.~\ref{fig:3specPhaseDiagram}a-c] are indicated as white lines in Fig.~\ref{fig:3specPhaseDiagram}d.

\subsection{Three-species coexistence}
Deviations from the two-species phase boundaries occur where nonlinearity dominates the population dynamics.
In particular, there is a small region of three-species coexistence in the overlap region of pairwise coexistence between $n_1$ and $n_2$, and $n_1$ and $n_3$, respectively (black region in Fig.~\ref{fig:3specPhaseDiagram}d).
Despite being small, this coexistence region covers a finite volume in high-dimensional parameter space.
Qualitatively, this small region can be understood from pairwise competition, too:
at the onset of pairwise $n_1$-$n_2$-coexistence (diamond symbol in Fig.~\ref{fig:3specPhaseDiagram}d), close to the region of inverted long-term dynamics, the dominant species is $n_2$ [Fig.~\ref{fig:3specPhaseDiagram}g].
In the same parameter region the overall gleaner $n_3$ goes extinct in $n_2$-$n_3$-competition, meaning that also here $n_2$ is the dominant species.
Thus, for three-species competition, there is $n_1$-$n_2$-coexistence as long as $n_2$ remains the dominant species in pairwise interactions.

Upon increasing the period duration $T$, approaching the triangle symbol in Fig.~\ref{fig:3specPhaseDiagram}d, $n_1$ takes over as the dominant species in pairwise coexistence with $n_2$.
$n_1$ can coexist with both $n_2$ and $n_3$ in this parameter region [Fig.~\ref{fig:3specPhaseDiagram}f].
Thus, a band of three-species coexistence emerges at this intermediate $T$.
Finally, upon further increasing the period duration $T$, $n_2$ and successively $n_3$ go extinct in pairwise competition with $n_1$.
In three-species competition, this corresponds to a region of $n_1$-$n_3$-coexistence, followed by regular long-term dynamics.

In the preceding discussions, we restricted the analysis to populations with identical decay rates $\delta$.
This fixes a global time scale for the population dynamics of all species.
However, these rates can be different when the decay is not dominated by uniform washout from a chemostat.
In this case, the time scales of the growth dynamics can be different for each species.
This leads to a more complex dependence of the pairwise coexistence regions on the period duration $T$ and can result in additional types of coexistence.
For example, Fig.~\ref{fig:3specPhaseDiagram}h shows a system where the time-dependent environment leads to pairwise coexistence between the overall gleaner and the overall opportunist, but leaves all other pairwise interactions untouched.
In the corresponding three-species ecosystem, the between-season population sizes $n_{i,k}$ vary slowly over the course of multiple periods ($T' \approx 120 T$) in addition to the within-season oscillating dynamics $n_i(t)$ with a periodicity $T$ determined by the external time dependence.
This highlights the impact of the ecosystem composition on the emerging temporal niches: in the specific example shown in Fig.~\ref{fig:3specPhaseDiagram}h there is a niche for species $n_1$ (yellow) as long as species $n_2$ (blue) is dominant, and similar relations can be found for the other species.
More species can be introduced to the ecosystem as long as they fit into this cycle of niches: a new species thriving only when $n_2$ is dominant but offering a niche to $n_1$ (and none of the other species) could coexist with all three species, and similar heuristic arguments can be made to further increase the number of species.
In Appendix~\ref{app:simulation} we demonstrate two cases of four-species coexistence.
However, as the number of species increases and between-season changes in the population sizes $n_{i,k}$ are induced, the minimal population size of individual species can vary over multiple orders of magnitude (e.g., $n_1$ in Fig.~\ref{fig:3specPhaseDiagram}h), and stochastic effects may become relevant~\cite{Ovaskainen.Meerson2010, Lande.etal2003}.

\section{Discussion and Conclusion}
In summary, we analyzed how coexistence between multiple species can be achieved in an explicitly time-dependent variant of the chemostat model.
We showed that the principle of competitive exclusion, which holds for time-independent models, does not necessarily apply anymore in the time-dependent variant.
Instead, the ecosystem can be deterred from reaching a steady state at which competitive exclusion applies, thus giving rise to interesting and rich population dynamics.
Our findings are consistent with previous research on the impact of temporal fluctuations on coexistence, in particular with resolutions of Hutchinson's ``paradox'' of phytoplankton biodiversity~\cite{Chesson1994,Hutchinson:1961,Litchman:2001tc,Barton.etal2010, Erez.etal2020}.

Our analysis of the population dynamics in a two-species chemostat system with periodically varying resource abundance shows that Tilman's R* rule can be generalized to time-dependent systems by calculating a steady-state population size from an approximate discretized map.
Importantly, we demonstrate that this map is sufficient to understand the qualitative population dynamics in a general time-dependent system.
In particular, we inferred from this analysis that there can be coexistence between two species when there is a balance of the advantageous periods for either species.
This is possible when the time scale at which the environment changes is comparable to the time scale at which the populations grow and, most importantly, requires a periodically varying environment.

Generalizing to multispecies systems , we demonstrated how complex population profiles in three-species systems can be deduced qualitatively from pairwise interactions.
This approach is consistent with recent experimental work on \textit{in vitro} ecosystems~\cite{Friedman:2017dk} and the \textit{C.~elegans} intestinal microbiome~\cite{Ortiz.etal2021}, where the composition of the full ecosystem can be inferred from the composition of a set of subsystems.
One of our key findings here was that a time-dependent environment can allow three-species or even four-species coexistence, even if -- as a consequence of competitive exclusion -- only one species can survive in a time-independent environment.

In this article, we limited our discussions to competing species models, with examples focused on chemostat systems.
In these systems, the interactions between the species often depend exclusively on the total population size.
In realistic ecosystems, however, populations can interact via many different mechanisms, such as sharing of multiple resources~\cite{Armstrong:1976}, the production of a common good~\cite{West.etal2006, Chuang.etal2009, Becker.etal2018, Cremer.etal2019}, or the production of a toxin~\cite{Czaran.etal2002, Weber.etal2014}.
In addition, populations in realistic ecosystem can be spatially structured, such that competition between species happens only at the boundaries of single-species communities~\cite{Reichenbach.etal2007, Dobramysl.etal2018}.
In agreement with previous work our results indicate that a time-dependent environment may be capable of enhancing the biodiversity in these systems, similar to the competing species model~\cite{White.Hastings2020, Barton.etal2010}.
However, a comprehensive analysis of the population dynamics in these systems in the presence of a time-dependent environment is still lacking.
Further extensions of the model discussed here could include the effect of demographic noise on the population dynamics, which is likely to play a decisive role in the highly nonlinear coexistence regime~\cite{Wienand.etal2017, Taitelbaum.etal2020}.

On a more general level, our results demonstrate the role of a time-dependent environment on the composition of an ecosystem.
In particular, our findings show that biodiversity in certain ecosystems can be enhanced by such a periodically varying environment or, equivalently, that biodiversity can be lost when removing an external fluctuation from the system.
This suggests that realistic ecosystems may depend crucially on natural environmental cycles, such as the circadian sunlight cycle or tidal ranges.
To test this hypothesis, experimental studies of ecosystems subject to temporal variations will be needed.

\section{Acknowledgements}
We thank Zheng Eelderink-Chen, Jeff Gore, Hyunseok Lee, Martha Merrow, and Uwe T\"auber for stimulating discussions and valuable input, and F.~Ra{\ss}hofer for critical reading of the manuscript.
We acknowledge financial support by the Deutsche Foschungsgemeinschaft through the Excellence Cluster ORIGINS under Germany’s Excellence Strategy (EXC-2094-390783311), and the funding initiative “What is life?” of the VolkswagenStiftung.
T.B.~acknowledges support by the Joachim Herz Foundation.

\appendix

\section{Representations of the chemostat model}\label{app:chemostatModel}%
In Eq.~\eqref{eq:chemostat} we use a chemostat model with an abiotic resource and assuming that all species consume the same amount of resources per capita.
Here we illustrate how to arrive at the proposed chemostat model from a more general version including a biotic resource $R_d(t)$ that is driven externally towards a target resource concentration $R(t)$, where the time dependence in $R(t)$ is externally imposed.
For brevity and without loss of generality we also omit the explicit enforcing of non-negative growth ($\text{max}(R_d(t)-N(t), 0) \to R_d(t) - N(t)$) by assuming that $R_d(t) > N(t)$ at all times.

The population dynamics in a generic consumer-resource model with a single resource $R_d(t)$~\cite{MacArthur1970, Murdoch.etal2013} read
\begin{equation}
    \frac{\mathrm{d}}{\mathrm{d} t} n_i(t) = n_i(t) \cdot \mu_i (R_d(t) - m_i)
\end{equation}
$\mu_i$ is the rate at which excess resources are consumed to produce offsprings and $m_i$ represents the amount of resources required to maintain a constant population size.
One may split $m_i$ into a term accounting for the total resources consumed by the entire ecosystem at maintenance $\sum_j q_j n_j(t)$, where $q_i$ is the resource quota representing the amount of resources consumed per capita~\cite{Grover:1997gr}, and a term representing the loss from death $L_i$.
The term $R_d(t) - \sum_j q_j n_j(t)$ then corresponds to the amount of excess resources available for reproduction, and the loss can be interpreted as a death rate $\delta_i = \mu_i L_i$.
Reproduction efficiency is linear in the resource excess at low resource density, but other factors should limit the reproduction speed at high resource density which is commonly accounted for via Monod-like saturation with a half-saturation constant $K_i$, resulting in the following set of equations for the population and resource dynamics:
\begin{subequations}\label{eq:app:consumer-resource}
\begin{align}
    \frac{\mathrm{d}}{\mathrm{d} t} n_i(t) &= n_i(t) \cdot \left( \mu_i \frac{R_d(t) - \sum_j q_j n_j(t)}{R_d(t) - \sum_j q_j n_j(t) + K_i} - \delta_i \right) \nonumber\\
    &= n_i(t) \left(\tilde \mu_i\bigl(R_d(t), \{n_j(t)\} \bigr) - \delta_i \right) \\
    \frac{\mathrm{d}}{\mathrm{d} t} R_d(t) &= r \, (R_d(t){-}R(t)) {-} \sum_i n_i(t) {\cdot} \tilde \mu_i\bigl(R_d(t), \{n_j(t)\} \bigr) 
\end{align}
\end{subequations}
where $\tilde \mu_i$ is a shorthand for the per-capita growth rate, and $r$ is the resource regulation rate.
Note that this per-capita growth rate is non-negative and therefore bound by $0 \leq \tilde \mu_i < \mu_i$, and that the maximum population size of each species is limited by the total number of resources $n_i(t) < R_d(t)/q_i$.
Further assuming without loss of generality that the maximum amount of resources in the system should be finite, $R_d(t) < R_{d,\text{max}}$ one can find an upper bound for the maximum resource consumption:
\begin{equation}
    \sum_i n_i(t) {\cdot} \tilde \mu_i\bigl(R_d(t), \{n_j(t)\} \bigr) < \sum_i \frac{R_{d,\text{max}}}{q_i} \, \mu_i = \text{const.}
\end{equation}
For resource regulation $r/R_{d,\text{max}}$ much faster than this resource consumption, the resource consumption can be neglected in Eq.~\eqref{eq:app:consumer-resource}, and the resource dynamics reduce to
\begin{equation}
    \frac{\mathrm{d}}{\mathrm{d} t} R_d(t) \approx r \, (R_d(t) - R(t)) \, .
\end{equation}
This equation corresponds to a biotic resource that decays towards a target resource level $R(t)$ at a rate $r$.
For sufficiently large $r$, this decay happens on a much shorter time scale than any change in the population sizes, and we can use a separation of time scales to arrive at an abiotic resource $R_d(t) \approx R(t)$ on the time scale of the population dynamics.

Next, one may map the population sizes $n_i$ to ``resource-consuming units'' by rescaling the population sizes by the resource quota~\cite{Grover:1997gr}, with $n_i \to n_i/q_i$, which together with the time scale separation above yields the chemostat model introduced in Eq.~\eqref{eq:chemostat}.
This model can be further rewritten by normalizing the population sizes in terms of the maximal resource concentration $n_i \to n_i \cdot R_\text{a}$, $K_i \to K_i \cdot R_\text{a}$, such that the abiotic resource is given by
\begin{equation}
    R(t) = 
\begin{cases}
	1 & \text{for } 0 \,{\leq}\, t \,{<}\, \nu T \, , \\
	R_\text{s}/R_\text{a} & \text{for } \nu T \,{\leq}\, t \,{<}\, T \, .
\end{cases}
\end{equation}
Similarly, one may rescale time in terms of one of the growth or death rates.
This is particularly appealing in the case of washout where all $\delta_i =\delta $ are identical, so that after rescaling $t \to t / \delta$, $T \to T/\delta$ and $\mu_i \to \mu_i \cdot \delta $ the chemostat model is
\begin{subequations}
\begin{align}
\frac{\mathrm{d} n_i(t)}{\mathrm{d} t}  &=
 n_i(t) \left[ \mu_i \, \frac{\max\left(R(t) {-} N(t),0\right)}{\max\left(R(t) {-} N(t),0\right) + K_i} - 1 \right], \\
R(t) &= 
\begin{cases}
	1 & \text{for } 0 \,{\leq}\, t \,{<}\, \nu T \, , \\
	R_\text{s}/R_\text{a} & \text{for } \nu T \,{\leq}\, t \,{<}\, T \, ,
\end{cases}
\end{align}
\end{subequations}
with $N(t) = \sum_i n_i(t)$.
In our analysis we omit the last steps of rescaling time and resource abundance since both $R_a$ and $\delta_i$ to highlight their role in the emergence of coexistence.

\section{Resource Buffer \texorpdfstring{$\K_i$}{K\_i}}
\label{app:resource_buffer}
The population dynamics of competing species as described in the main matter follows the chemostat model
\begin{equation}
 \label{eq:app:chemostat}
 \frac{\mathrm{d}}{\mathrm{d} t} n_i(t) =
 n_i(t) \cdot \left( \mu_i \frac{R - N(t)}{R - N(t) + K_i} - \delta_i \right).
\end{equation}
Consider a system inhabited by a single species, so that $N(t) = n_i(t)$.
The differential equation~\ref{eq:app:chemostat} has two fixed points, $n_i(t)=0$ and $n_i(t) = R-\K_i =: \bar N_i$, with $\K_i$ given by
\begin{equation}
 \K_i = \frac{\delta_i}{\mu_i-\delta_i} \, K_i \, .
\end{equation}
This offset denotes the amount of resources that are left unbond by the species $i$ upon reaching its non-zero steady state.
This is a direct effect of the resource-limited Monod-like growth.
In a system hosting two species, the resources left unbound by species $i$ are available for species $j$.
However, species $j$ can only feed on them if its own resource buffer $\K_j$ is not reached yet, i.e.~if $\K_j<\K_i$.
This is the R* rule.
\par
If the time dependence does not affect the amount of resources, but any of the other system parameters, then the resource buffer may become time-dependent itself, 
\begin{equation}
 \K_i(t) = \K_i(\mu_i(t),\, \delta_i(t),\, K_i(t)) \, .
\end{equation}
Similar to the case of time-dependent resources abundance, this may lead to coexistence and inversion, but also allows for temporal niches~\cite{Levins:1979tu,Armstrong:1976,Chesson:2000,Nowack:2018}.

\section{Constraints on the growth function \texorpdfstring{$f_i$}{f\_i}}\label{app:growth_function}
In the general growth model class of competing species models, the population dynamics are governed by a growth function  $f_i(N(t),\,t)$~\cite{Hirsch:1982so}:
\begin{equation}
 \frac{\mathrm{d} n_i(t)}{\mathrm{d} t} = n_i(t) \cdot f_i(N(t),\,t) \, .
\end{equation}
Between two species, the growth functions $f_i$ can in general differ at certain parameters (e.g.,~different growth rates $\mu_i$), or the $f_i$ may be completely different functions of the parameters.
For a growth function to be realistic and biologically meaningful, it needs to meet several requirements: 
(i)~The concept of \emph{competing} species is incorporated in this function by requiring that -- for any species -- the growth should be slower if any population size increases (while all others remain constant).
(ii)~In an almost abandoned environment ($N \to 0$), growth should always be possible.
(iii)~There should be a single threshold population size $\n_i$ at which the population cannot grow anymore, and above which the growth is negative, as the system can only sustain a limited number of individuals.
Mathematically, the conditions (i)-(iii) can be expressed as
\begin{subequations}
\begin{align}
 \text{(i)} && \frac{\partial}{\partial n_j} \, f_i(N(t),\,t) &< 0 \, , \\
 \text{(ii)} &&  f_i(N\to 0,\,t) &> 0 \, , \\
 \text{(iii)} &&  f_i\left(N > \n_i,\,t\right) &< 0 \, .
\end{align}
\end{subequations}
In the chemostat model defined in Eq.~\eqref{eq:chemostat}, the threshold population size is identical to the carrying capacity, $\n_i = \bar N_i$.
In addition, to ensure continuous dynamics, the growth functions $f_i$ need be finite for all states that are accessible in reality, i.e.~for all $\{n_j{>}0\}$.

\section{Derivation of the discrete map}\label{app:map}
\begin{figure}
    \centering
    \includegraphics[width = \columnwidth]{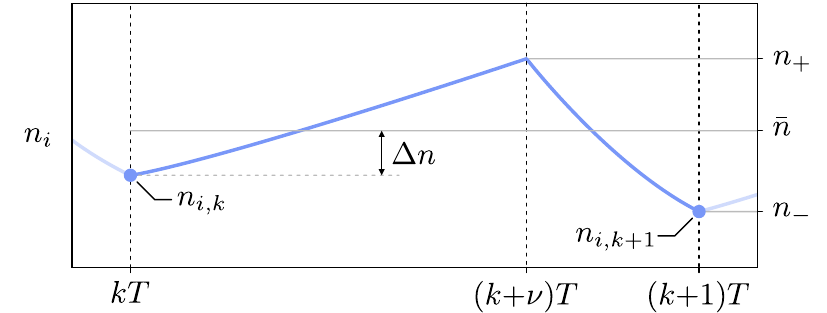}
    \caption{(Color online)
    Conceptual time evolution of the population size of a single species over one period.
    For the entire period, the population size remains within an interval bounded by $n_+$ and $n_-$.
    Note that in the chemostat model as defined in Eq.~\eqref{eq:app:chemostat}, the minimal population size $n_-$ coincides with $n_{i,k{+}1}$, which need not be the case in general.}
    \label{fig:app:map}
\end{figure}
In competing species models, the population dynamics are determined by the growth functions $f_i$.
The functional form of the growth functions is determined by the system under investigation.
The wide variety of ecosystems and models makes it difficult to make general statements about the population dynamics, or let alone solve the corresponding differential equations analytically.
In addition, a common feature of most population dynamics models, and in particular of the competing species models, is the nonlinearity of the differential equations that represent the population dynamics, which further complicates the analysis.
However, the nonlinearity can be neglected to lowest order when propagating the system over short time intervals for any competing species model, as we will show in the following.

Consider a system hosting one species only.
To further simplify the explanations, we begin by defining two new quantities: over the course of one oscillation, the population size will reach a local maximum $n_+$ and a local minimum $n_-$ [Fig.~\ref{fig:app:map}].
We will demonstrate in the following that the error introduced by the approximate map is to lowest order proportional to $T \cdot f_i$ within a single period.
To see this, formally solve the differential equation~\ref{eq:competition-model} for a single species, i.e., for the case where $N(t) = n_i(t)$:
\begin{equation}
    \label{eq:app:exact_solution}
    n_i(t{+}T) = n_i(t) \cdot \exp\left[\int\limits_t^{t+T} \mathrm d u \, f_i\bigl(n_i(u),\, u\bigr) \right] \,.
\end{equation}
Making use of the competition condition in Eq.~\eqref{eq:competing-condition}, stating that $\partial_{n_i} f_i \leq 0$, it follows that the growth function in the integral is bounded from above and below by $n_+$ and $n_-$.
In addition, for any arbitrary external time dependence, there is always one $u_+$ ($u_-$) that maximizes (minimizes) the growth function at a constant population size, so that
\begin{equation}
    f_i (n_+,\, u_-) < f_i(n_i(u),\,u) < f_i(n_-,\, u_+) \quad \forall u \in [t,\, t{+}T]\, .
\end{equation}
This allows to calculate upper and lower bounds to the integral in Eq.~\eqref{eq:app:exact_solution}, and thus also for the population size after one period:
\begin{equation}
    n_-\, e^{f_i (n_+,\, u_-) T} < n_i(t+T) < n_+\, e^{f_i (n_-,\, u_+) T} \,.
\end{equation}
Expanding the exponential to lowest order in $T$ and using crude approximations for $n_- > n_i(t) \cdot \exp\bigl[f_i(n_+, u_-)\bigr]$  and $n_+ < n_i(t) \cdot \exp\bigl[f_i(n_-, u_+)\bigr]$, this shows that the population varies within a range proportional to $T$ over a single period.
In addition, from the mean value theorem it follows that there exists a constant $\bar n$ with $n_- {<} \bar n {<} n_+$ [Fig.~\ref{fig:app:map}] so that
\begin{equation}
    \label{eq:app:exact_solution_discrete}
     n_i(t{+}T) = \bar n \cdot \exp\left[\int\limits_t^{t+T} \mathrm d u \, f_i\bigl(\bar n,\, u\bigr) \right] \,.
\end{equation}
This $\bar n$ differs from the population size at the beginning of the period $n_{i,k}$ by $\bar n = n_{i,k} + \Delta n$.
Notably, from the definition of $\bar n$ it follows immediately that $\Delta n < n_+ - n_- \sim \mathcal O (T \cdot f_i)$.
Thus, expanding the growth function around $\bar n$, we find
\begin{equation}
    f_i (n_{i,k}, u) = f_i(\bar n, u) + \partial_{n_i} f(\bar n_i, u) \, \Delta n + \mathcal O(\Delta n^2) \,.
\end{equation}
Substituting this into Eq.~\eqref{eq:app:exact_solution_discrete} and renaming $n_i(t+T)$ to $n_{i,k+1}$ leads -- to lowest order in $T \cdot f_i$ -- to the discrete map from Eq.~\eqref{eq:map}:
\begin{align}
    n_{i,k+1} &=  n_{i,k} \cdot \exp\left[\int\limits_t^{t+T} \mathrm d u \, f_i\bigl(n_{i,k},\, u\bigr) \right] + \mathcal O(\Delta n) \nonumber \\
    &= n_{i,k} \cdot \exp\left[\int\limits_t^{t+T} \mathrm d u \, f_i\bigl(n_{i,k},\, u\bigr) \right] + \mathcal O(T \cdot f_i) \, .
\end{align}
The same argument holds for a system hosting more than one species.
In this case, the upper and lower limits $n_\pm$ as well as the constant $\bar n$ are replaced by a set of corresponding quantities.
The decisive observation, namely that $\Delta n_i < n_{i,+} - n_{i,-} \sim \mathcal O(T \cdot f_i)$, remains valid for any number of species.
Thus, the discrete map Eq.~\eqref{eq:map} approximates the exact dynamics up to $\mathcal O(T \cdot f_i)$, making it reasonable for period durations short compared to the time scales of growth.
\section{Derivation of phase boundaries}\label{app:phase_boundaries}
In a nonlinear competing species model hosting two species, the time-dependent environment can lead to an inversion of the long-term population dynamics compared to a time-independent environment, and possibly coexistence.
Such inversion and coexistence can only be established for a range of parameters.
For two given species in an environment that switches between to given distinct states, the parameters that quantify the periodic structure are the duration of one period $T$ and the fraction of one period $\nu$ that is spent in one of the two states.
Note that for other time-dependent systems, for example an environment that changes continuously, there may be other parameters that quantify the temporal structure.
In the following, we will explain how the phase boundaries to inverted long-term dynamics and to coexistence can be obtained for a general system of two competing species.
Along the lines, we will discuss in detail how the phase boundaries on $\nu$ and $T$ are obtained for the chemostat model with varying resources.
Throughout this discussion, the species are labeled such that $n_1$ is the gleaner and $n_2$ is the opportunist ($\K_1 < \K_2$).

\subsection{Formal derivation}
For the purpose of determining the biodiversity of an ecosystem, we distinguish between four distinct states of the system:
\begin{enumerate*}[label=(\roman*)]
 \item the gleaner survives,
 \item the opportunist survives,
 \item both populations survive, or
 \item neither population survives. 
\end{enumerate*}
In a time-independent environment, the asymptotic state of the system can be obtained by performing a linear stability analysis on the fixed points $\n_i$ of the ODEs modeling the population dynamics.
However, this is not possible in a time-dependent environment, since the fluctuation of model parameters prevents the system from reaching a steady state.
Thus, the asymptotic dynamics of such a system are periodic trajectories, $\n_i(t)$, rather than fixed points.
In a nonlinear system, these trajectories can in general only be determined exactly by numerically solving the ODEs, which yields little information about the conceptual dynamics.
By approximating the asymptotic trajectories, however, the phase boundaries can be estimated.

In a two-species system, the two asymptotic trajectories where only one of the two populations, $n_i$, survives while the other is extinct, respectively, are of particular relevance.
Formally, the stability of these asymptotic trajectories with respect to invasion by the other species $n_j$ can be assessed from the average net growth rate
\begin{equation}
    \langle f_j \rangle (t) = \frac{1}{T} \int\limits_t^{t+T} \mathrm d u \, f_j(\{n^*_i(u)\},\,u) \, ,
\end{equation}
where $\n_i(u) \approx N(t)$ is the asymptotic trajectory of the prevailing species and $n_j(u) \approx 0$ is negligibly small.
By comparing this to Eq.~\eqref{eq:app:exact_solution}, it is obvious that $n_{j,k{+}1} = n_{j,k} \exp[T \cdot \langle f_j (k T)\rangle ]$.
Thus, if the average net growth rate is positive (negative), the invading population grows (goes extinct) and the asymptotic trajectory of the prevailing species is unstable (stable).
Hence, phase boundaries are located at parameter combinations where any average net growth rate changes sign.

\begin{figure*}[!t]
 \scriptsize
 \includegraphics[width = \textwidth]{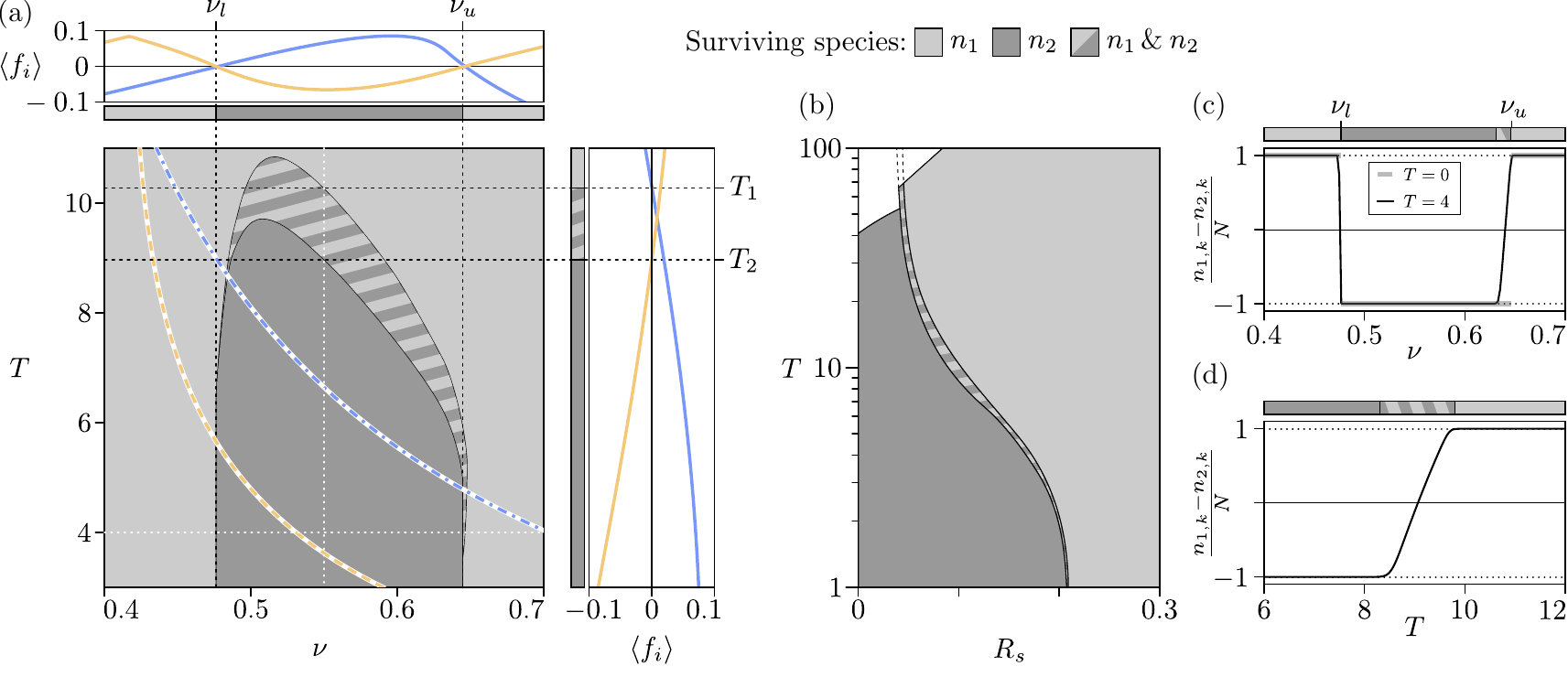}
 \caption{\label{app:fig:invasion}(Color online) (a) Simulated net growth rates $\langle f_i \rangle$ for the gleaner ($n_1$, yellow or light gray) and the opportunist ($n_2$, blue or dark gray), along two representative cutlines (dotted white) in the phase diagram of two-species competition.
 Bars next to the growth rate plots show the predicted outcome of the competition, based on the net growth rates: each species can survive if the net growth rate is positive, $\langle f_i \rangle >0$.
 When both net growth rates are positive simultaneously, coexistence at a stable equilibrium is predicted.
 Lines in the phase diagram show the estimated threshold period durations $\widetilde T_1(\nu)$ (yellow, dashed) and $\widetilde T_2(\nu)$ (blue, dash-dotted), providing results accurate within an order of magnitude.
 (b) Bifurcation diagram showing the transition from $R_\text{s} = 0$ (corresponding to panel (a) and Fig.~\ref{fig:phase_space}b) to $R_\text{s} > 0$ (corresponding to Fig.~\ref{fig:phase_space}c) for $\nu = 0.55$.
 The white area indicates parameter regions where both species are considered extinct as described in Appendix~\ref{app:simulation}. Dashed lines are extrapolations of the boundaries for the coexistence region.
 (c) Relative population size differences $(n_1-n_2)/N$ at $ t = 100\,T$ with $T = 4$. The solid line indicates stable equilibria, while dotted lines indicate unstable equilibria. For comparison, the stable equilibria at $T\to 0$ are shown as light gray lines. Boxes: the regions of regular (light gray) and inverse (dark gray) long-term dynamics are separated by narrow regions of coexistence.
 (d) Relative population size differences at $ t = 100\,T$ with $\nu = 0.55$. The coexistence region is bounded by two transcritical bifurcations. In the coexistence region, the dominant species changes smoothly from $n_2$ to $n_1$ with increasing $T$.}
\end{figure*}%

\subsection{Phase boundary on Inversion}
To derive the phase boundaries to the region of inverted long-term dynamics, consider a system with a short period duration $T$ compared to the time scales off growth, $1/f_i$.
As discussed in Appendix~\ref{app:map}, one may accurately approximate the exact population dynamics in the limit $T \ll 1/f_i$ by a discrete map
\begin{equation}
    n_{i,k+1} = n_{i,k} \cdot \exp\left[ \int\limits_0^T \mathrm d t \, f_i\left(N_k,\,t\right) \right] \, .
\end{equation}
In particular, this map can be used to test whether one population can invade the other population.
Inverted long-term dynamics means that the opportunist species $n_2$ can invade a residing gleaner population $n_1$.
In the following, we denote the steady state population size of the gleaner as $\n_1$, such that $n_{1,k} = \n_1$ for all $k$ at the steady state.
By definition, this steady state corresponds to the carrying capacity $\bar N_1$.
The opportunist can invade the gleaner population if the average net growth within one period is positive,
\begin{equation}
    \langle f_2 \rangle \left (\nu) \right|_{N = \n_1} := \frac{1}{T} \int\limits_0^T \mathrm d t \, f_2(\n_1,\, t) >0 \, .
\end{equation}
Otherwise, the opportunist population will go extinct.
By definition, this net growth is negative for time-independent environments, $\nu=0$ and $\nu = 1$, and varies continuously when changing the activity ratio $\nu$.
Thus, if the net growth is positive for any value of $\nu$, this implies that there are two threshold values for the activity ratio, $\nu_u$ and $\nu_l$, at which $\langle f_2\rangle (\nu_{u,l}) = 0$.
These threshold values mark the phase boundaries of the region of inverted long-term dynamics at short period durations (Fig.~\ref{app:fig:invasion}a).

Notably, the gleaner steady state population size depends on the activity ratio, too, $\n_1(\nu)$, and hence also $\bar N_1(\nu)$.
In particular, the gleaner species may not be able to survive even without competition for some activity ratios, such that $\n_1 = 0$.
This is the case, for example, in the chemostat model (Eq.~\eqref{eq:td-chemostat}) for the lower boundary on the inversion region.
For arbitrary $\nu$, the discrete map can be used to formally obtain the gleaner steady state population size at $R_\text{s}=0$ for $T \ll 1/f_i$:
\begin{align}
    \label{eq:app:steady_state}
    0 &= \int\limits_t^{t+T} \mathrm du \, f_1(\n_1(\nu),\,u) \,, \nonumber \\
   \rightarrow \qquad \n_1(\nu) &\equiv \bar N_1(\nu) = R_\text{a} - K_1 \frac{\delta_1}{\mu_1 \nu - \delta_1} \,.
\end{align}
However, as only non-negative steady state population sizes are meaningful, this steady state population size will be zero when
$$ \nu < \frac{\delta_1}{\mu_1} \cdot \frac{K_1+R_\text{a}}{R_\text{a}} =: \nu_1 \, .$$
A similar equation can be derived for the steady state population size of the opportunist.
Thus, the opportunist can survive in an environment in which the gleaner goes extinct for ${\nu_1 > \nu > \nu_2}$.
In particular, the smallest activity ratio for which the average net growth for the opportunist is positive is $\nu_2$, so that the lower bound to the region of inverted long-term dynamics is $\nu_l = \nu_2$ [Fig.~\ref{fig:phase_space}a].

The upper bound to the phase of inverted long-term dynamics in the chemostat model can be calculated straightforwardly by solving $\langle f_2 \rangle(\nu_u) = 0$ using the expression for $\n_1(\nu)$ derived in Eq.~\eqref{eq:app:steady_state} in place of the total population size.
Hence, the phase boundaries of inverted long-term dynamics for short period durations $T$ are given by
\begin{subequations}
\begin{align}
    \nu_l &= \frac{\delta_2}{\mu_2} \cdot \frac{K_2 + R_\text{a}}{R_\text{a}} \,,\\
    \nu_u &= \frac{\delta_1 \delta_2 (K_1 - K_2)}{\mu_2 \delta_1 K_1 - \mu_1 \delta_2 K_2} \, .
\end{align}
\end{subequations}

The example above corresponds to the special case of zero resources during the period with resources absent in the chemostat model, $R_\text{s}=0$.
The same arguments hold true for any other competing species model, in particular for the chemostat model with $R_\text{s}>0$.
This method was used to calculate the phase boundaries at short period durations $T$ in Figs.~\ref{fig:phase_space} and~\ref{fig:3specPhaseDiagram}a-c.

\subsection{Phase boundary on Coexistence}
The results above were derived for the case where the period duration $T$ is short compared to the time scales of growth, $1/f_i$.
This allowed to circumvent the nonlinearity of the dynamics by using a discrete map in order to analyse the characteristic population dynamics.
However, it is precisely these nonlinear dynamics that lead to competitive exclusion in a time-independent system.
Similarly, as the period duration $T$ in a system with externally imposed time dependence becomes long enough such that a steady state is reached before the environment switches, the opportunist's short-term advantage vanishes and the conventional R* rule can come into effect.

\begin{figure}
 \scriptsize
 \includegraphics[width = \columnwidth]{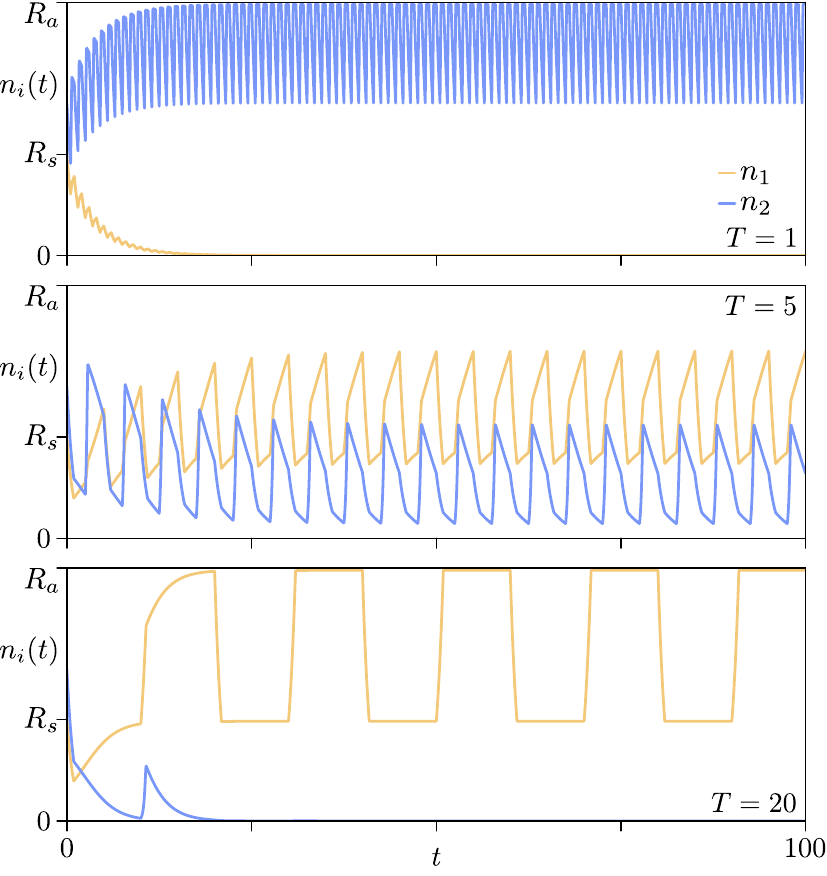}
 \caption{\label{app:fig:varyT}(Color online) Simulated trajectories of two-species competition at different period durations $T$ showing the transition from regular ($T=20$) to inverted ($T=1$) dynamics, with a band of coexistence ($T=5$) in between. Parameters are specified in Table~\ref{tab:params}.}
\end{figure}

At what period duration $T$ will this transition from inverted to regular long-term dynamics occur?
Inverted long-term dynamics occur as long as the gleaner cannot invade a prevailing opportunist population, i.e., as long as the average net growth rate
\begin{equation}
    \langle f_1 \rangle \left (\nu) \right|_{N = \n_2(t)} = \frac{1}{T} \int\limits_0^T \mathrm d t \, f_1(\n_2(t),\, t) <0 \, ,
\end{equation}
where $\n_2(t)$ is the asymptotic trajectory of the opportunist.
Thus, a change in the qualitative dynamics occurs at a period duration $T_1$ at which $\left.\langle f_1 \rangle  \right|_{T_1}=0$.
Similarly, a threshold value $T_2$ can be obtained above which the opportunist cannot invade a gleaner population.
For intermediate period durations $T_1 < T < T_2$, either species can invade the other, so that coexistence between these species is possible [Figs.~\ref{app:fig:invasion},~\ref{app:fig:varyT}].

However, since the asymptotic trajectories are in general not known, these phase boundaries cannot be calculated exactly.
Instead, they can be estimated by approximating the asymptotic trajectories, or from the qualitative dynamics of the system.
For the chemostat model as defined in Eq.~\eqref{eq:td-model}, for example, the nonlinearities leading to coexistence between two species become relevant when the total population size approaches the steady state population size $R(t)-\K_i$ of the dominant species.
Thus, for any period duration $T$ at which the total population size gets close to $R(t)-\K_i$ before the environment changes, we expect the nonlinear interactions to be sufficiently relevant to favor the gleaner population.

To estimate the corresponding period duration $T$ for each species independently, assume that the population size at the beginning of each period $t = k\cdot T$ is $\left. \n_i \right|_{R_\text{s}>0}$, as obtained from Eq.~\eqref{eq:app:steady_state} for $R_\text{s}> 0$.
We furthermore assume the growth dynamics to be exponential, thereby ignoring the nonlinearities for this estimate.
The period duration $\widetilde T$ at which the population reaches the steady state population size $R_\text{a}-\K_i$ is used as an estimate for the threshold period duration:
\begin{align}
    \label{eq:app:threshold_period_duration}
   R_\text{a} - \K_i &= \left. \n_i \right|_{R_\text{s}} \cdot \exp \left[ \int\limits_0^{\nu \widetilde T_i} \mathrm d t \, f_i\left(\left. \n_i \right|_{R_\text{s}}, t \right) \right] \nonumber \, , \\
    \rightarrow \quad \widetilde T_i(\nu) &= \frac{1}{\nu} \, \log \left( \frac{R_\text{a}-\K_i}{\left. \n_i \right|_{R_\text{s}}} \right) \cdot \frac{1}{f_i(\left. \n_i \right|_{R_\text{s}}, t )} \, .
\end{align}
Importantly, this estimate confirms that the inverted long-term dynamics can be expected only for $T \ll 1/f_i$ at non-zero $R_\text{s}$.
Note that this threshold period duration depends on the scarcity resource level $R_\text{s}$ indirectly through $\left. \n_i \right|_{R_\text{s}}$.
As $R_\text{s} \to 0$, this population size at the beginning of each period tends to zero, and $\tilde T_i$ diverges, implying that the band of coexistence moves towards higher values of the period duration $T$ [Fig.~\ref{app:fig:invasion}b].

\section{Numerical solution of the ODE}\label{app:simulation}
\begin{figure}
 \scriptsize
 \includegraphics[width = \columnwidth]{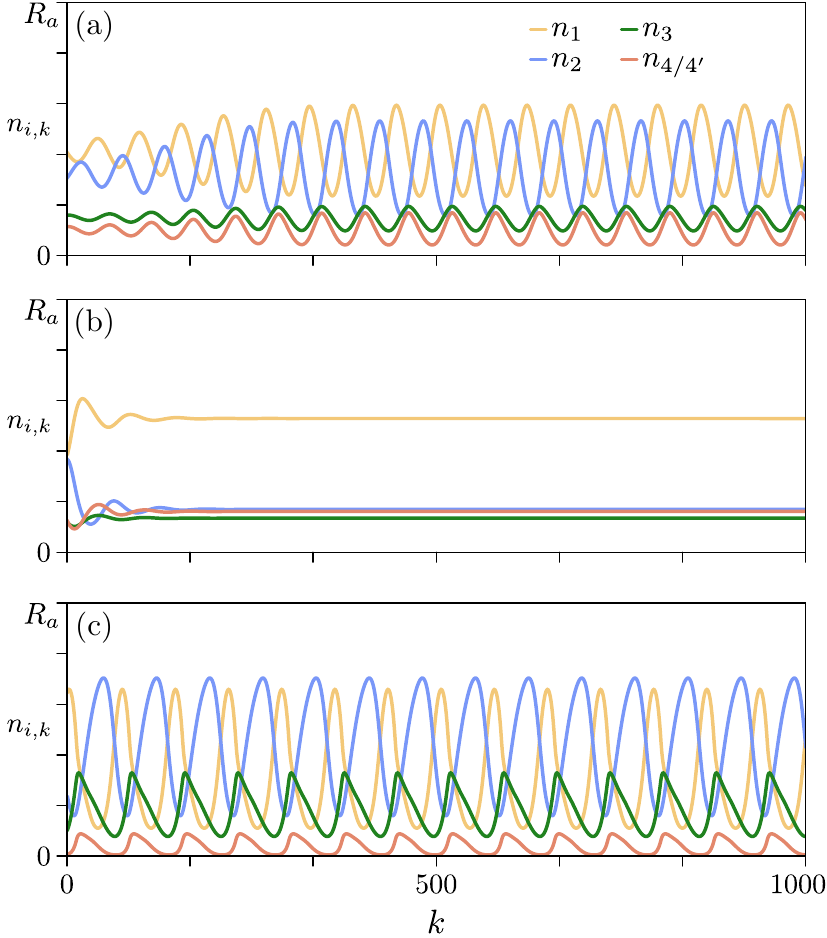}
 \caption{\label{app:fig:4spec}(Color online) Sample trajectories of four-species competition with parameters $\mu_4=1.4$, $\delta_4=0.2$, and $K_4=0.13$, for two different initial conditions, (a) showing limit cycle dynamics and (b) approaching a fixed point of the between-season dynamics. (c) For a different parameter set ($\mu_{4'}=1.4$, $\delta_{4'}=0.2$, $K_{4'}=0.13$) the limit cycle is independent of the initial conditions. All parameters are specified in Table~\ref{tab:params}.}
\end{figure}

\begin{table}[!t]
    \centering
\begin{tabular}{l | l|r|| l|r|| l|r|}
    \cline{2-7}
     \multirow{3}{*}{Figs.~2-5, 8\hspace{1em}} & $\mu_1$ \hspace{.6em} & \hspace{.6em} $2.0$ & $\delta_1$ \hspace{.6em} & \hspace{.6em} $1.0$ & $K_1$ \hspace{.6em} & \hspace{.6em} $0.2\phantom{00}$\vphantom{\large M}\\
      & $\mu_2$ & $6.0$ & $\delta_2$ & $1.0$ & $K_2$ & $2.0\phantom{00}$ \\
      \cline{2-7}
      & $R_\text{a}$ & $1.0$ & $R_\text{s}$ & $ 0.1$ & \multicolumn{2}{c}{\vphantom{\large M}}\\
      \cline{2-5}
      \multicolumn{7}{c}{\vspace{-.8em}}\\
      \cline{2-7}
      \multirow{4}{*}{Fig.~6a-g\hspace{1em}} & $\mu_1$ & $1.8$ & $\delta_1$ & $1.0$ & $K_1$ & $0.003$\vphantom{\large M}\\
      & $\mu_2$ & $2.7$ & $\delta_2$ & $1.0$ & $K_2$ & $0.018$ \\
      & $\mu_3$ & $4.5$ & $\delta_3$ & $1.0$ & $K_3$ & $0.090$ \\
      \cline{2-7}
      & $R_\text{a}$ & $1.0$ & $R_\text{s}$ & $ 0.6$ &\multicolumn{2}{c}{\vphantom{\large M}}\\
      \cline{2-5}
      \multicolumn{7}{c}{\vspace{-.8em}}\\
      \cline{2-7}
      \multirow{5}{*}{Figs.~6h, 10\hspace{1em}} & $\mu_1$ & $0.9$ & $\delta_1$ & $0.5$ & $K_1$ & $0.003$\vphantom{\large M}\\
      & $\mu_2$ & $1.1$ & $\delta_2$ & $0.4$ & $K_2$ & $0.012$ \\
      & $\mu_3$ & $0.7$ & $\delta_3$ & $0.1$ & $K_3$ & $0.130$ \\
      & $\mu_4$ & $1.4$ & $\delta_4$ & $0.2$ & $K_4$ & $0.130$ \\
      & $\mu_{4'}$ & $1.6$ & $\delta_{4'}$ & $0.3$ & $K_{4'}$ & $0.066$ \\
      \cline{2-7}
      & $R_\text{a}$ & $1.0$ & $R_\text{s}$ & $ 0.4$ &\multicolumn{2}{c}{\vphantom{\large M}}\\
      & $T$ & $10$ & $\nu$ & $ 0.7$ &\multicolumn{2}{c}{\vphantom{\large M}}\\
      \cline{2-5}
      \multicolumn{7}{c}{\vspace{-.8em}}\\
      \cline{2-7}
     \multirow{3}{*}{Fig.~9\hspace{1em}} & $\mu_1$ \hspace{.6em} & \hspace{.6em} $2.0$ & $\delta_1$ \hspace{.6em} & \hspace{.6em} $1.0$ & $K_1$ \hspace{.6em} & \hspace{.6em} $0.003$\vphantom{\large M}\\
      & $\mu_2$ & $6.0$ & $\delta_2$ & $1.0$ & $K_2$ & $0.033$ \\
      \cline{2-7}
      & $R_\text{a}$ & $1.0$ & $R_\text{s}$ & $ 0.4$ &  $\nu$ & $ 0.5$\vphantom{\large M} \\
      \cline{2-7}
\end{tabular}
    \caption{Parameters used for numerically solving the ODEs and generating the plots.}
    \label{tab:params}
\end{table}%
To study the population dynamics in a specific system and to generate the figures, the differential equations in Eq.~\eqref{eq:td-chemostat} were solved numerically.
Unless specified otherwise, the parameters stated in Table~\ref{tab:params} were used throughout all numerical solutions.
These parameters were chosen such that they fulfill the requirements for inverted long-term dynamics (two-species competition) and three-species coexistence (three-species competition) for a wide range of parameters $\nu$ and $T$, and with easily discernible visual features.
For four-species competition, the parameter set from three-species competition was extended by another species with $K_4=K_3$ and twice as fast growth and decay ($\mu_3 = 2\, \mu_4$ and $\delta_3=2 \,\delta_4$), where we observe that limit cycle of the population dynamics depends on the initial conditions [Fig.~\ref{app:fig:4spec}a,b].
For a fourth species that is unrelated to all others ($\mu_{4'}$, $\delta_{4'}$, $K_{4'}$) the limit cycle is independent of the initial conditions [Fig.~\ref{app:fig:4spec}c].

All phase diagrams were obtained by solving the differential equations numerically for each parameter combination $(\nu,\,T)$ separately.
The numerical solver was terminated when one of the following criteria was met:
\begin{enumerate*}[label=(\roman*)]
 \item The population size of either species dropped below a threshold of $R_\text{a} \cdot 10^{-30}$ at the end of one period, i.e.\ at $t= k\cdot T$.
 \item A hard time limit of $t=10^{3} \cdot T$ was exceeded.
\end{enumerate*}

In order to classify the results, the population sizes at the final time step ($n_{i,k_\text{max}}$) as well as the corresponding logarithmic population size change ($\log(n_{i,k_\text{max}}-n_{i,k_\text{max}-1})/T$) were used to bin a parameter combination as ``regular long-term dynamics", ``inverted long-term dynamics" or ``coexistence".
A species was considered extinct if the population size was below the threshold of $n_{i,k_\text{max}} < R_\text{a} \cdot 10^{-30}$.
If both populations remained above this threshold until the solver stopped, the logarithmic population size change was used as a secondary criterion: a species was considered as extinct if $\log(n_{i,k_\text{max}}-n_{i,k_\text{max}-1})/T < -10^{-4}$.
This threshold was chosen since it well predicted the outcome observed for larger $T$.
If parameter combination was tagged as ``regular long-term dynamics" if only the gleaner survived and ``inverted long-term dynamics" if only the opportunist survived.
If both populations survived, the parameter combination was tagged as ``coexistence".

For the invasion plots in Fig.~\ref{app:fig:invasion}, the system was simulated for one period with initial conditions $n_i(0) = 10^{-5}$, $n_j(0) = n_j^*(0) - n_i(0)$.
Here, $n_i$ is the invading population and $n_j$ is the residing population, with its fixed trajectory $n_j^*(t)$.
The net population change of the invading species after one period was used to determine the net growth rate, $\langle f_i \rangle = \log(n_i(T)/n_i(0))/T$.

\section{Pairwise competition}\label{app:pairwise}
In many-species ecosystems, an analytical study of the population dynamics is often difficult owed to the high-dimensionality and nonlinearity of the system.
For competing species models where the interactions depend only on the total population size, however, it is sufficient to study pairwise interactions between species in order to understand the qualitative time evolution of the entire ecosystem.
In the following, we justify this hypothesis.

Consider a general competing species model as defined in Eq.~\eqref{eq:general-model},
\begin{subequations}
\begin{align}
&\frac{\mathrm{d}}{\mathrm{d} t} n_i(t)  =
n_i(t) \cdot  f_i(N(t) \, , \\
&\frac{\mathrm{\partial}}{\mathrm{\partial} n_j} f_i(N)  \leq
 0 \, ,
\end{align}
\end{subequations}
in a system hosting $M$ distinct species.
For simplicity, we assume a time-independent environment in the following, so that there is no explicit time dependence in the growth functions.
Assume furthermore that this system is predominantly inhabited by one species, $n_1$, with all other species contributing only marginally to the total population size, $n_1 \gg n_i$ with $i \in \{2,\ldots,M\}$.
Now, consider the general case where the total population size is far from the dominant species' steady state population size $\n_1$.
Then, since $\partial_t n_i(t) \sim n_i(t)$, the nonlinear growth ensures that the population size of $n_1$ can vary much more quickly than all other population sizes, $\partial_t n_1(t) \gg \partial_t n_i(t)$.
Thus, the dominant species will quickly approach its steady state population size.
Since this species contributes primarily to the total population size, the total population size will vary from this steady state population size only by a small amount, $N(t) = \n_1 + \Delta N(t)$.
This allows to estimate the population dynamics for all species to lowest order:
\begin{subequations}
\label{eq:app:total-pop-size}
\begin{align}
    \frac{\mathrm d}{\mathrm d t} n_1(t) &= n_1(t) \cdot f_1(\n_1 + \Delta N(t)) \nonumber \\
    &= n_1(t) \cdot \partial_n f_1(N)\left. \right|_{\n_1} \, \Delta N(t) + \mathcal O(\Delta N^2) \, , \label{eq:app:dominant}\\
    \frac{\mathrm d}{\mathrm d t} n_i(t) &= n_i(t) \cdot f_i(\n_1 + \Delta N(t)) \nonumber \\
    &= n_i(t) \cdot f_i(\n_1) + \mathcal O (n_i \cdot \Delta N) \, . \label{eq:app:subdominant}
\end{align}
\end{subequations}
Notably, since $\partial_N f_i \leq 0$ in the competing species model, the dominant population size changes such that the deviation $\Delta N$ from the steady state population size is minimized.
This ensures that the total population size remains close to $N(t) \approx \n_1$ irrespective of the population changes of all other species, as long as $n_1$ is the dominant species.
This shows that the growth rates of all subdominant populations is to lowest order determined by the steady state population size of the dominant species, $\n_1$.
Thus, in a many-species system, the population dynamics of all subdominant species are equivalent to a two-species system where the dominant species is the only competitor.
In other words, the pairwise competition with the dominant species is sufficient to characterize the dynamics of all subdominant populations.

This reasoning can be extended to systems with an external time dependence.
In these systems, a single population does not approach a fixed point, but rather an asymptotic trajectory $\n_i(t)$.
Following the same line of arguments, the total population size in such systems predominantly inhabited by one species $n_1(t) \gg n_i(t)$ is approximated by the asymptotic trajectory of this dominant species, $N(t) \approx \n_1(t)$.
This means that also in a time-dependent environment it is sufficient to consider pairwise competition to characterize the entire system's population dynamics.

\vfill

\bibliography{arxiv_v3.bib}

\end{document}